\documentclass[11pt,a4paper,dvipsnames]{article}

\usepackage{CJK}

\usepackage{tikz}
\usetikzlibrary{calc,matrix,decorations.pathmorphing,decorations.markings,arrows,positioning,intersections,mindmap,backgrounds,patterns,arrows.meta,chains}

\usepackage{graphbox}

\usepackage{jstyle}
\usepackage{amsmath, amsfonts, amssymb}
\usepackage{mathtools, extarrows}
\usepackage{bbm}
\usepackage{graphicx, hyperref, verbatim}
\usepackage{xcolor}
\usepackage{float}
\usepackage{subfig}
\usepackage{caption}

\DeclareGraphicsExtensions{.pdf,.png,.jpg,.mps}
\usepackage{booktabs}
\usepackage{multirow}
\usepackage{cancel}
\usepackage{array}
\newcolumntype{P}[1]{>{\centering\arraybackslash}p{#1}}
\newcolumntype{M}[1]{>{\centering\arraybackslash}m{#1}}

\usepackage{physics}
\usepackage{diagbox}
\usepackage{pifont}
\usepackage{cleveref}
\crefname{equation}{}{}

\usepackage{genyoungtabtikz}

\usepackage{lmodern} 
\usepackage{scalerel,accents,trimclip}

\usepackage{fancybox}
\usepackage{tcolorbox}

\usetikzlibrary{calc}

\title{Spinning Mellin amplitudes}

\author[a]{Zhongjie Huang (黄中杰)}
\author[b,c]{Yichao Tang (唐一朝)}

\affiliation[a]{Kavli Institute for Theoretical Sciences, University of Chinese Academy of Sciences, Beijing 100190, China}
\affiliation[b]{Institute of Theoretical Physics, Chinese Academy of Sciences, Beijing 100190, China}
\affiliation[c]{School of Physical Sciences, University of Chinese Academy of Sciences, Beijing 100049, China}

\emailAdd{huangzhongjie@ucas.ac.cn}
\emailAdd{chaostang@outlook.com}

\abstract{We propose a definition of Mellin amplitudes for conformal correlators involving arbitrary spinning operators in tensor representations of the Lorentz group. These representations cover all bosonic local operators. Our strategy is to perform discrete Mellin transforms on all scalar products involving polarization vectors, so that each polarization vector can be interpreted as the position of a fictitious scalar operator. We establish the general pole structures and factorization properties of these spinning Mellin amplitudes. We also provide a systematic algorithm to derive factorization formulas with arbitrary spinning exchanges, yielding new explicit results up to spin-4. To illustrate the practicality of our formalism, we bootstrap the 3- and 4-point current correlators in a 4d $\mathcal{N}=2$ superconformal field theory, which are dual to gluon scattering amplitudes in $\mathrm{AdS}_5 \times \mathrm{S}^3$. The results agree with the snowflake channel of $6$- and $8$-point scalar supergluon amplitudes in the literature. In particular, we prove that at tree level, there is no $F^3$ interaction in the effective bulk Lagrangian.}

\begin{document} 

\begin{CJK*}{UTF8}{}
\CJKfamily{gbsn}
\maketitle
\end{CJK*}

\tableofcontents
	
\newpage

\section{Introduction}

Since its introduction by Mack \cite{Mack:2009mi}, the Mellin formalism has proven to be a natural language for studying correlation functions in conformal field theories (CFTs), particularly those with holographic duals \cite{Penedones:2010ue,Fitzpatrick:2011ia,Fitzpatrick:2011hu}. In this formalism, correlation functions transform into Mellin amplitudes, which usually exhibit simpler analytic structures compared to the position space functions: operators exchanged in the operator product expansion (OPE) appear as poles in Mellin amplitudes, and the residues factorize neatly into products of sub-amplitudes \cite{Goncalves:2014rfa}. When a CFT has a large central charge and is dual to a perturbative quantum field theory in anti-de Sitter space (AdS), exchanged operators (and thus the poles in Mellin amplitudes) correspond to particles propagating in the AdS bulk \cite{Heemskerk:2009pn,El-Showk:2011yvt}, and Mellin amplitudes act as scattering amplitudes in AdS. Such similarity to flat-space scattering amplitudes makes the Mellin formalism especially suitable for the analytic bootstrap of holographic correlators, yielding profound results over the past decade (see \cite{Bissi:2022mrs,Heslop:2022xgp} for reviews). Recent studies of non-perturbative Mellin amplitudes \cite{Penedones:2019tng} have further revealed deep connections between the Lorentzian inversion formula \cite{Caron-Huot:2017vep,Simmons-Duffin:2017nub} and Mellin space dispersion relations \cite{Caron-Huot:2020adz}. 

Despite its success, the Mellin formalism remains mostly limited to scalar correlators, largely due to the absence of a definition for Mellin amplitudes of generic spinning correlators. Solving this long-standing problem is crucial for studying scattering processes in AdS, as many particles of interest, such as gluons and gravitons, carry spin. Motivation also comes from recent research on five- and higher-point functions \cite{Goncalves:2019znr,Alday:2022lkk,Goncalves:2023oyx,Alday:2023kfm,Cao:2023cwa,Cao:2024bky,Huang:2024dxr,Goncalves:2025jcg,Fernandes:2025eqe}. To fully exploit the factorization properties of Mellin amplitudes, it is sometimes necessary to consider multi-channel factorization, which generally involves sub-amplitudes with multiple spinning operators and requires a consistent definition of spinning Mellin amplitudes.

In this paper, we propose a definition for Mellin amplitudes with arbitrary spinning operators in tensor representations of the Lorentz group, which include all bosonic local operators. We mostly focus on traceless symmetric tensors for simplicity. The basic idea is straightforward: spinning operators can be described by their position $X^M$ and polarization $Z^M$ in embedding space \cite{Weinberg:2010fx,Costa:2011mg,Simmons-Duffin:2012juh}:
\begin{align}
    \mathcal O(X,Z)=Z^{M_1}\cdots Z^{M_J}\mathcal O_{M_1\cdots M_J}(X),
\end{align}
and the correlation functions $\langle\mathcal O(X_1,Z_1)\cdots\mathcal O(X_N,Z_N)\rangle$ depend on three kinds of scalar products $X_p \cdot X_q$, $Z_p \cdot X_q$, and $Z_p \cdot Z_q$\footnote{We work in generic spacetime dimension $d$ and restrict our discussion to the parity-even sector. Parity-odd structures involving the Levi-Civita tensor $\epsilon$ are beyond the scope of this paper. }. In the conventional definition of scalar Mellin amplitudes, $X_p \cdot X_q$ is associated with the Mellin variables $\gamma_{pq}$ via a (continuous) Mellin transform. For spinning Mellin amplitudes, we simply perform additional discrete Mellin transforms on the new scalar products $Z_p \cdot X_q$ and $Z_p \cdot Z_q$:
\begin{equation}
    \begin{aligned}
        \langle\mathcal O(X_1,Z_1)\cdots\mathcal O(X_N,Z_N)\rangle=\sum_{\eta\in\mathbb Z}\int&[{\rm d}\gamma]\, \mathcal M(\gamma_{pq},\eta_{p'q},\eta_{p'q'}) \prod_{\substack{p,q=1\\p<q}}^N\frac{\Gamma(\gamma_{pq})}{(-2X_p\cdot X_q)^{\gamma_{pq}}} \\
        & \times \prod_{\substack{p,q=1\\p\neq q}}^N\frac{(2Z_p\cdot X_q)^{\eta_{p'q}}}{\eta_{p'q}!}\prod_{\substack{p,q=1\\p<q}}^N\frac{(2Z_p\cdot Z_q)^{\eta_{p'q'}}}{\eta_{p'q'}!}.
    \end{aligned}
\end{equation}
The Mellin amplitudes now depend on two kinds of Mellin variables\footnote{Similar discrete Mellin transforms have been used to deal with the R-symmetry structures in the $\rm AdS\times S$ formalism \cite{Aprile:2020luw}.}, the continuous $\gamma$ and the discrete $\eta$. The primes in the indices indicate which labels are tied to the polarization $Z$. 

The key advantage of our definition is that it treats all scalar products $X\cdot X$, $Z\cdot X$, and $Z\cdot Z$ on an equal footing. As a result, the factorization properties of spinning Mellin amplitudes follow directly from those of scalar Mellin amplitudes. A simple way to see this is that the continuous and discrete Mellin transforms are related through an identity \eqref{eq:similar} (and \eqref{eq:analyticity}), which allows us to ``Wick rotate'' all discrete variables into continuous ones, and treat $\eta=-\gamma$ when analyzing factorization. From this perspective, the polarization $Z_p$ can be viewed as the position $X_{p'}$ of a fictitious scalar operator, and the factorization formula for $N$-point spinning Mellin amplitudes then follows directly from the factorization formula of $2N$-point scalar Mellin amplitudes. The above argument is confirmed by a detailed investigation of the OPE and the factorization properties of these spinning Mellin amplitudes. 

We remark that there have been attempts to define spinning Mellin amplitudes in the literature\footnote{See also \cite{Faller:2017hyt} for Mellin amplitudes involving fermionic operators.}. The first approach \cite{Paulos:2011ie,Chu:2023pea,Chu:2023kpe} keeps all tensor indices manifest. This makes it easier to translate position space Feynman rules of Witten diagrams to Mellin space, but the resulting Mellin amplitude is cluttered with indices, which complicates the computation. The second approach \cite{Chen:2017xdz,Sleight:2018epi}, restricted to four-point correlators, chooses to perform a discrete Mellin transform on the independent tensor structures $V_{i,jk}$ and $H_{ij}$ involving polarization. While these structures manifest the embedding-space property $\mathcal O(X,Z+\beta X)=\mathcal O(X,Z)$, they mix $Z$ and $X$ in a nontrivial manner. As a result, spinning Mellin amplitudes defined in this way have significantly more complicated factorization properties, which makes the computation impractical in higher-point generalizations. Our definition avoids these shortcomings, treating spinning and scalar operators on an equal footing and manifesting simple factorization properties at any multiplicity.

Finally, we emphasize that although we mostly focus on traceless symmetric tensors, our definition naturally extends to operators with arbitrary mixed-symmetry tensors. Such operators can be described by introducing additional polarization vectors, and the corresponding Mellin amplitude is defined by performing Mellin transforms over all scalar products. An operator described by $h$ polarization vectors effectively corresponds to $h+1$ scalar operators in the Mellin amplitude.

This paper is organized as follows. In section~\ref{sec:def}, we introduce our definition of spinning Mellin amplitudes. In section~\ref{sec:opepole}, we relate the multi-particle OPE to the pole structure of spinning Mellin amplitudes, generalizing the argument in~\cite{Fitzpatrick:2011ia}. In section~\ref{sec:factorization}, we discuss the general strategy to obtain factorization formulas by solving a Casimir equation recursively. In section~\ref{sec:boots}, we demonstrate the practical utility of our definition by bootstrapping the three-point gluon amplitudes in ${\rm AdS}_{d+1}$ dual to generic CFTs as well as some four-point gluon amplitudes in $\mathrm{AdS}_5\times \mathrm{S}^3$ in a specific $\mathcal N=2$ superconformal field theory (SCFT). Finally, in Section \ref{sec:outlook}, we conclude and discuss some future directions.

\section{Definition of spinning Mellin amplitudes}\label{sec:def}

\subsection{Spinning operators and the embedding formalism}\label{subsec:embedding}

We consider CFTs in dimension $d\geq3$ and work in Euclidean signature. In this signature, the metric is given by $g_{\mu\nu}=\delta_{\mu\nu}$, and the corresponding conformal group is $SO(d+1,1)$. Local operators in CFTs are classified by their conformal dimensions $\Delta$ and their irreducible representations of the Lorentz group $SO(d)$. The simplest cases are scalar operators $\mathcal{O}(x)$, which transform trivially under rotations. Other representations carry non-trivial spins and can be written as tensors\footnote{We only consider bosonic operators in this paper. In general, spinors are needed for fermionic operators.} whose indices satisfy certain symmetries. These tensors can be described by Young diagrams. The most common spinning operators are rank-$J$ traceless symmetric tensors $O_{\mu_1\cdots\mu_J}(x)$ (``spin-$J$'' operators), which correspond to single-row Young diagrams with $J$ boxes:
\begin{equation}
\gyoung(^2<\longleftarrow\ >^4J^2<\ \longrightarrow>,;_6<\cdots>;)
\end{equation}
To keep track of all the components of these tensors, we can write them in index-free notation~\cite{Bargmann:1977gy} using a null polarization vector $z^\mu$:
\begin{equation}
\mathcal O(x,z)=z^{\mu_1}\cdots z^{\mu_J}\mathcal O_{\mu_1\cdots\mu_J}(x),\quad z^2=0,
\end{equation}
so that the symmetry property is automatic and the tracelessness is ensured by the null condition of the polarization.

In general, all irreducible tensor representations of $SO(d)$ can be written as traceless tensors with mixed symmetries. These operators correspond to multi-row Young diagrams and carry multiple spin labels. For example, the two-row Young diagram (``spin-$(J_{(1)},J_{(2)})$'' operators)
\begin{equation}
\gyoung(^2<\longleftarrow\ >^4<J_{(1)}>^2<\ \longrightarrow>,;_6<\cdots>;,;_4<\cdots>;,^2<\longleftarrow\ >^2<J_{(2)}>^2<\ \longrightarrow>)
\end{equation}
implies that $\mathcal O_{\mu_1\cdots\mu_{J_{(1)}},\,\nu_1\cdots\nu_{J_{(2)}}}(x)$ is first symmetrized under $\mu_i\leftrightarrow\mu_j$, $\nu_i\leftrightarrow\nu_j$ and then anti-symmetrized under $\mu_i\leftrightarrow\nu_i$. The index-free notation~\cite{Costa:2016hju} now assigns different polarizations for each row:
\begin{equation}
\mathcal O(x,z_{(1)},z_{(2)})=\mathcal O_{\mu_1\cdots\mu_{J_{(1)}},\,\nu_1\cdots\nu_{J_{(2)}}}z_{(1)}^{\mu_1}\cdots z_{(1)}^{\mu_{J_{(1)}}}z_{(2)}^{\nu_1}\cdots z_{(2)}^{\nu_{J_{(2)}}},
\end{equation}
with
\begin{equation}
    z_{(1)}^2=z_{(2)}^2=z_{(1)}\cdot z_{(2)}=0.
\end{equation}
The anti-symmetry under $\mu_i\leftrightarrow\nu_i$ implies that replacing any one of the $z_{(2)}$ with $z_{(1)}$ will yield 0, which is equivalent to the property
\begin{equation}
z_{(1)}\cdot\frac\partial{\partial z_{(2)}}\mathcal O(x,z_{(1)},z_{(2)})=0\quad\implies\quad\mathcal O(x,z_{(1)},z_{(2)}+\beta z_{(1)})=\mathcal O(x,z_{(1)},z_{(2)}).
\end{equation}
Similarly, for Young diagrams with $h$ rows, we have operators that depend on $h$ polarizations:
\begin{equation}
\mathcal O(x,z_{(1)},\cdots,z_{(h)}),\quad z_{(i)}^2 = z_{(i)}\cdot z_{(j)} = 0,
\end{equation}
and satisfy
\begin{equation}
    \mathcal O(x,\cdots,z_{(i)} + \sum_{j<i} \beta_{ij}z_{(j)},\cdots) = \mathcal O(x,\cdots,z_{(i)},\cdots).
\end{equation}

Using the embedding formalism~\cite{Weinberg:2010fx,Costa:2011mg,Simmons-Duffin:2012juh}, we can describe these spinning operators and their correlation functions in the embedding space $\mathbb R^{d+1,1}$, where the conformal group is linearly realized as $SO(d+1,1)$ rotations. The spacetime coordinate $x^\mu$ is lifted to a projective null ray $X^M\sim \lambda X^M$ passing through the origin in $\mathbb R^{d+1,1}$:
\begin{equation}
    X^M=(X^0,X^\mu,X^{d+1}),\quad X\cdot X= -(X^0)^2+(X^1)^2+\cdots+(X^{d+1})^2 =0,
\end{equation}
so the CFT and all the operators are located on the light-cone $X\cdot X=0$. We can return to the $d$-dimensional space by taking the Poincar\'e section
\begin{equation}\label{eq:poincare}
    (X^+,X^\mu,X^-)=(1,x^\mu,x\cdot x),\quad\text{where } X^\pm := X^0\pm X^{d+1}.
\end{equation}
The spinning operators are promoted to transverse tensors on the light-cone, and the polarizations $z_{(i)}^\mu$ are similarly lifted to $Z_{(i)}^M$, satisfying
\begin{equation}
    Z_{(i)}\cdot Z_{(i)} = Z_{(i)}\cdot Z_{(j)} = 0,\quad Z_{(i)}\cdot X = 0
\end{equation}
due to the tracelessness and transversality. These polarizations are related to $z_{(i)}^\mu$ through the parameterization
\begin{equation}
    (Z_{(i)}^+,Z_{(i)}^\mu,Z_{(i)}^-)=(0,z_{(i)}^\mu,2x\cdot z_{(i)}),\quad\text{where } Z_{(i)}^\pm := Z_{(i)}^0\pm Z_{(i)}^{d+1}.
\end{equation}
The uplift of operators from $d$-dimensional space to the embedding space can be achieved as follows:
\begin{equation}\label{eq:spinOp}
    \mathcal O(X,Z_{(1)},\cdots,Z_{(h)})=(X^+)^{-\Delta}\mathcal O\left(x^\mu=\frac{X^\mu}{X^+},\cdots,z_{(i)}^\mu=Z_{(i)}^\mu-\frac{Z_{(i)}^+}{X^+}X^\mu,\cdots\right).
\end{equation}
These spinning operators in the embedding space now satisfy the following equations 
\begin{align}
    \mathcal O(\lambda X,\cdots,\alpha_iZ_{(i)},\cdots)=&\ \lambda^{-\Delta}\alpha_i^{J_i}\mathcal O(X,\cdots,Z_{(i)},\cdots),\label{eq:mixweight}\\
    \mathcal O(X,\cdots,Z_{(i)}+\beta X,\cdots)=&\ \mathcal O(X,\cdots,Z_{(i)},\cdots),\\
    \mathcal O(X,\cdots,Z_{(i)}+\sum_{j<i}\beta_{ij}Z_{(j)},\cdots)=&\ \mathcal O(X,\cdots,Z_{(i)},\cdots).
\end{align}
Note that the last two equations can be combined into one as 
\begin{align}
    \mathcal O(X,\cdots,Z_{(i)}+\beta_{i0}X+\sum_{j<i}\beta_{ij}Z_{(j)},\cdots)=\mathcal O(X,\cdots,Z_{(i)},\cdots).\label{eq:mixgauge}
\end{align}
This indicates that the position $X$ of the operators can be viewed as a kind of ``polarization'' $Z_{(0)}$ with spin $-\Delta$ in the embedding space, and vice versa. This idea serves as the core of our definition of spinning Mellin amplitudes. Later, we will make this statement more precise. In particular, for the $p$-th operator $\mathcal O(X_p,Z_{p,(1)},\cdots,Z_{p,(h)})$ in the correlation function, we would like to denote 
\begin{subequations}\label{eq:prime1}
    \begin{align}
        X_{p'}:=Z_{p,(1)},\quad \Delta_{p'} := -J_{p,(1)},\\
        X_{p''}:=Z_{p,(2)},\quad \Delta_{p''} := -J_{p,(2)},
    \end{align}
\end{subequations}
and similarly for the other polarizations and spin labels. 

We will mainly focus our discussion on the traceless symmetric operators, where there is only one polarization $Z$. Explicitly, these operators satisfy
\begin{gather}
    \mathcal O(\lambda X,\alpha Z)=\lambda^{-\Delta}\alpha^J\mathcal O(X,Z),\label{eq:confweight}\\
    \mathcal O(X,Z+\beta X)=\mathcal O(X,Z)\ \Longleftrightarrow\ (X\cdot\partial_Z)\mathcal O(X,Z)=0.\label{eq:confgauge}
\end{gather}
For symmetric traceless operators we fix the two-point normalization by
\begin{equation}\label{eq:twoPointNormalization}
    \langle\mathcal O(X_1,Z_1)\mathcal O(X_2,Z_2)\rangle
    =\frac{\big[(2X_1\cdot Z_2)(2X_2\cdot Z_1)-(2X_1\cdot X_2)(2Z_1\cdot Z_2)\big]^J}{(-2X_1\cdot X_2)^{\Delta+J}}.
\end{equation}
Our operators obey $\mathcal O=2^{J/2}\mathcal O_{\text{\cite{Goncalves:2014rfa}}}$, since uniformly attaching a factor of $2$ to every scalar product is the most natural choice in our framework.

\subsection{Scalar Mellin amplitudes}

The Mellin representation~\cite{Mack:2009mi,Penedones:2010ue,Fitzpatrick:2011ia} provides a nice representation for scalar correlators that manifests conformal invariance. The Mellin amplitude $\mathcal M(\gamma)$ is defined as the Mellin transform of the (connected) correlation function:
\begin{equation}\label{eq:continuousMellin}
    G(X_1,\cdots,X_N)=\int[{\rm d}\gamma]\,\mathcal M(\gamma)\prod_{\substack{p,q=1\\p<q}}^N\frac{\Gamma(\gamma_{pq})}{(-X_{pq})^{\gamma_{pq}}},
\end{equation}
where $X_{pq}:=2X_p\cdot X_q$. The Mellin variables $\gamma_{pq}$ are symmetric in their indices and satisfy the constraints
\begin{equation}\label{eq:gammaconstraints}
    \forall\,1\leq p\leq N:\quad\sum_{\substack{q=1\\q\neq p}}^N\gamma_{pq}=\Delta_p.
\end{equation}
As a result, there are $\frac12N(N-3)$ independent Mellin variables, and the integration in \eqref{eq:continuousMellin} is over any set of independent Mellin variables. The freedom of choosing independent Mellin variables allows us to write the amplitudes in many different ways. Therefore, it is useful to regard Mellin amplitudes as functions defined on the support of \eqref{eq:gammaconstraints}. It is also very helpful to think of the Mellin variables as products of auxiliary momenta $\mathtt k_p$ that satisfy momentum conservation and on-shell conditions:
\begin{equation}
    \gamma_{pq}=\mathtt k_p\cdot\mathtt k_q,\quad \sum_{p=1}^N\mathtt k_p=0,\quad\mathtt k_p^2=-\Delta_p.
\end{equation}
This notation automatically solves \eqref{eq:gammaconstraints}. For this reason, we also call \eqref{eq:gammaconstraints} the ``momentum conservation'' constraints. These auxiliary momenta offer further insight into the pole structures of Mellin amplitudes, which we discuss below.

The Mellin amplitude $\mathcal M(\gamma)$ is a meromorphic function\footnote{We restrict to perturbative Mellin amplitudes in this paper. In a non-perturbative CFT, the location of poles in $\mathcal M(\gamma)$ could have accumulation points due to the large-spin operators~\cite{Penedones:2019tng}, spoiling meromorphicity at these accumulation points.} with poles corresponding to operators exchanged in the OPE. We will discuss this in more detail in section~\ref{sec:opepole}. For now, we simply note that the OPE channel
\begin{equation}\label{eq:OPEscalar}
    \langle\mathcal O(X_1)\cdots\mathcal O(X_\ell)|\mathcal O_{\Delta,J}|\mathcal O(X_{\ell+1})\cdots\mathcal O(X_N)\rangle
\end{equation}
for the exchange of an operator $\mathcal O_{\Delta,J}$ with dimension $\Delta$ and spin $J$ corresponds to a sequence of poles in the Mellin amplitude~\cite{Goncalves:2014rfa}
\begin{equation}
    \mathcal M(\gamma)\sim\frac{\mathcal Q_m}{\gamma_{LR}-\tau-2m},\quad m=0,1,2,\cdots,\quad\tau:=\Delta-J.
\end{equation}
Here,
\begin{equation}
    \gamma_{LR}:=\sum_{a\in\mathcal L}\sum_{i\in\mathcal R}\gamma_{ai}=\sum_{a\in\mathcal L}\mathtt k_a\cdot\sum_{i\in\mathcal R}\mathtt k_i
\end{equation}
can be interpreted as the Mandelstam invariant of the auxiliary momentum flow across the OPE channel, where the sets $\mathcal L=\{1,\cdots,\ell\}$ and $\mathcal R=\{\ell+1,\cdots,N\}$ contain all external labels on the left and right, respectively. In the following, we will always use $a,b,\cdots$ and $i,j,\cdots$ to denote external labels on the left and right, and omit the range specifications $a,b\in\mathcal L$ and $i,j\in\mathcal R$ in the sums. The residues $\mathcal Q_m$ are given by the product of the left and right half-amplitudes $\mathcal M_{\mathcal L}$ and $\mathcal M_{\mathcal R}$ under certain shifts. For example, if the exchanged operator is a scalar ($J=0$), we have (see (87) in \cite{Goncalves:2014rfa})
\begin{equation}\label{eq:facdemoscalar}
    \mathcal Q_m = \frac{-2\Gamma(\Delta)m!}{(1+\Delta-\frac{d}{2})_m}\left(\frac{1}{2^m m!}  \hat{\mathbbm x}^m  \circ \mathcal M_{\mathcal L}\right) \left(\frac{1}{2^m m!}  \hat{\mathbbm x}^m\circ \mathcal M_{\mathcal R}\right),
\end{equation}
where $(1+\Delta-\frac{d}{2})_m$ is the Pochhammer symbol, defined as
\begin{equation}
    (\gamma)_m:=\gamma(\gamma+1)\cdots(\gamma+m-1).
\end{equation}
The shift operator $\hat{\mathbbm x}$ is defined by\footnote{Strictly speaking, this formula can only be used after we have substituted the Mellin variables involving the exchanged operator in terms of $\gamma_{ab}$ and $\gamma_{ij}$ in $\mathcal M_{\mathcal L/\mathcal R}$ using momentum conservation. We will discuss this subtle issue in section~\ref{sec:factorization}.}
\begin{equation}\label{eq:xhat}
    \hat{\mathbbm x} \circ \mathcal M_{\mathcal L}=\sum_{\substack{a,b\\a\neq b}}\gamma_{ab}[\mathcal M_{\mathcal L}]^{ab},\quad 
        \hat{\mathbbm x} \circ \mathcal M_{\mathcal R}=\sum_{\substack{i,j\\i\neq j}}\gamma_{ij} [\mathcal M_{\mathcal R}]^{ij},
\end{equation}
where we have introduced the abbreviation  
\begin{equation}\label{eq:shift}
    [\mathcal M]^{pq}:=\mathcal M(\gamma_{pq}\mapsto\gamma_{pq}+1),\quad [\mathcal M]_{pq}:=\mathcal M(\gamma_{pq}\mapsto\gamma_{pq}-1).
\end{equation}
Explicit formulas for the exchange of spin-1 and spin-2 operators can also be found in~\cite{Goncalves:2014rfa}.

In general, the residue $\mathcal Q_m$ can be determined by studying the Casimir equation for the specific OPE channel \eqref{eq:OPEscalar} in Mellin space. Consider the Casimir operator $\mathcal C$ acting on the left half of the external operators:
\begin{equation}
    \mathcal C:=\frac12\left[\sum_a\left(X_{aM}\frac{\partial}{\partial X_a^N}-X_{aN}\frac{\partial}{\partial X_a^M}\right)\right]^2.
\end{equation}
In Mellin space, it becomes a difference operator $\hat{\mathbb C}$ \cite{Fitzpatrick:2011ia,Goncalves:2014rfa} that acts on the amplitude as
\begin{equation}\label{eq:casMelscalar}
    \hat{\mathbb C}\mathcal M=\gamma_{LR}(d-\gamma_{LR})\mathcal M+\sum_{\substack{a,b\\a\neq b}}\sum_{\substack{i,j\\i\neq j}}\left(\gamma_{ai}\gamma_{bj}(\mathcal M-[\mathcal M]_{aj,bi}^{ai,bj})+\gamma_{ab}\gamma_{ij}[\mathcal M]_{ai,bj}^{ab,ij}\right).
\end{equation}
As in the derivation of conformal blocks, the Casimir acting on the left external operators can be replaced by its action on the exchanged operator. This gives rise to the Casimir equation of the OPE channel \eqref{eq:OPEscalar} in Mellin space
\begin{equation}\label{eq:casPos}
    \hat{\mathbb C}\mathcal M=-c_{\Delta,J}\mathcal M,\quad c_{\Delta,J}:=-\Delta(d-\Delta)+J(d-2+J),
\end{equation}
and the corresponding residues $\mathcal Q_m$ can then be solved recursively from this equation. We will discuss how to solve the Casimir equation in detail in section~\ref{sec:factorization}.

\subsection{Discrete Mellin transform}\label{subsec:discrete}

Before we move on to the definition of spinning Mellin amplitudes, let us introduce the important notion of the discrete Mellin transform. In correlation functions of spinning operators, one encounters not only scalar products $X_p \cdot X_q$, but also terms like $Z_p \cdot X_q$ and $Z_p \cdot Z_q$. The dependence of the correlator on the polarizations is always polynomial. Such polynomials do not admit a well-defined Mellin amplitude under \eqref{eq:continuousMellin}. To deal with polynomials we must instead work with a discrete version of the Mellin transform.

Let $F(Z_1,\ldots,Z_N)$ be a polynomial constructed from products $Z_p \cdot Z_q$, with scaling behavior $F \sim Z_p^{J_p}$ for each $p$. Its discrete Mellin transform is defined by
\begin{equation}\label{eq:discreteMellin}
    F(Z_1,\cdots,Z_N)=\sum_{\eta\in\mathbb Z}\mathcal M(\eta)\prod_{\substack{p,q=1\\p<q}}^{N}\frac{(Z_{pq})^{\eta_{p'q'}}}{\eta_{p'q'}!},
\end{equation}
where $Z_{pq}:=2Z_p \cdot Z_q$. The primes on the labels in $\eta$ indicate that these variables are the Mellin transform of $Z$. This notation will prove convenient when we have Mellin transforms involving scalar products of both $X$ and $Z$. The scaling of $F$ imposes the constraints 
\begin{equation}\label{eq:mc}
    \forall\,1\leq p\leq N:\quad\sum_{\substack{q=1\\q\neq p}}^N\eta_{p'q'}=J_p,
\end{equation}
and the Mellin amplitudes are defined on the support of the above constraints. 

When performing the sum in \eqref{eq:discreteMellin}, the factorials $\eta_{p'q'}!$ in the denominators force $\eta_{p'q'}$ to be non-negative integers (for negative values, $\eta_{p'q'}!$ diverges, so $1/\eta_{p'q'}!$ vanishes). The constraints \eqref{eq:mc} therefore reduce the summation range to a finite number of lattice points. As a result, \eqref{eq:discreteMellin} becomes a finite sum and always produces a polynomial. It is also important to note that only the values of $\mathcal M(\eta)$ at these finitely many lattice points are meaningful. Consequently, there is an additional level of ambiguity in presenting the discrete Mellin amplitude, apart from the ``momentum conservation'' ambiguity of \eqref{eq:mc}. For example, consider the polynomial
\begin{equation}\label{eq:soExample}
    F(Z_1,Z_2,Z_3,Z_4)=Z_{13}Z_{24}+Z_{14}Z_{23}-Z_{12}Z_{34}.
\end{equation}
Taking the amplitude to be 
\begin{equation}\label{eq:proposal1}
    \mathcal M(\eta)=\eta_{1'3'}\eta_{2'4'}+\eta_{1'4'}\eta_{2'3'}-\eta_{1'2'}\eta_{3'4'}
\end{equation}
correctly reproduces $F$ after summing over all the contributing lattice points
\begin{equation}\label{eq:latticeExample}
    (\eta_{1'2'},\eta_{1'3'},\eta_{1'4'},\eta_{2'3'},\eta_{2'4'},\eta_{3'4'})=(1,0,0,0,0,1),\ (0,1,0,0,1,0),\ (0,0,1,1,0,0).
\end{equation}
However, the alternative form
\begin{equation}\label{eq:proposal2}
    \mathcal M(\eta)=\frac{1-(-1)^{\eta_{1'3'}}}2\eta_{2'4'}+42\,\eta_{1'3'}\eta_{1'4'}+\frac{\eta_{1'4'}\eta_{2'3'}}{\eta_{1'2'}+1}-\frac12\eta_{1'2'}\eta_{3'4'}(\eta_{3'4'}+1)
\end{equation}
evaluates to the same values at the lattice points in \eqref{eq:latticeExample} and thus reproduces the same polynomial. Both presentations are equivalent, though clearly the former is more natural than the latter. In subsection~\ref{subsec:singlespin}, we will impose further restrictions on the form of $\mathcal M$ to make it more physically interpretable.

Although the kernel in the discrete Mellin transform looks quite different from that in \eqref{eq:continuousMellin}, they are in fact related by the simple identity
\begin{equation}\label{eq:similar}
    \mathop{\rm Res}_{\gamma=-\eta}\frac{\Gamma(\gamma)}{(-P)^\gamma}=\frac{(+P)^\eta}{\eta!},\quad\eta=0,1,2,\cdots.
\end{equation}
This inspires us to adopt the notation 
\begin{equation}\label{eq:prime2}
    \gamma_{p'q}:=-\eta_{p'q},\quad\gamma_{p'q'}:=-\eta_{p'q'}
\end{equation}
to write the continuous and discrete Mellin variables in the same manner. This notation will not cause any confusion, since the $\gamma$'s with primed indices should always be understood as discrete Mellin variables. We will use the two symbols $\gamma$ and $-\eta$ interchangeably in the rest of the paper. 

\subsection{Spinning Mellin amplitudes}\label{subsec:definition}

With both continuous and discrete Mellin transforms, we can now present the definition of spinning Mellin amplitudes. For simplicity, we first focus on traceless symmetric operators. The definition for mixed symmetry operators will be given at the end of this subsection. For a (connected) spinning correlation function, we define
\begin{equation}\label{eq:spinMellin}
    \begin{aligned}
        &G(X_1,Z_1,\cdots,X_N,Z_N)\\
        =\,&\sum_{\eta\in\mathbb Z}\int[{\rm d}\gamma]\,\mathcal M(\gamma,\eta)\prod_{\substack{p,q=1\\p<q}}^N\frac{\Gamma(\gamma_{pq})}{(-2X_p\cdot X_q)^{\gamma_{pq}}}\prod_{\substack{p,q=1\\p\neq q}}^N\frac{(2Z_p\cdot X_q)^{\eta_{p'q}}}{\eta_{p'q}!}\prod_{\substack{p,q=1\\p<q}}^N\frac{(2Z_p\cdot Z_q)^{\eta_{p'q'}}}{\eta_{p'q'}!},
    \end{aligned}
\end{equation}
which is the Mellin transform, continuous or discrete, for all the scalar products. The Mellin variables $\gamma_{pq},\eta_{p'q},\eta_{p'q'}$ are symmetric in their indices and satisfy the following constraints due to the weight counting \eqref{eq:confweight}:
\begin{equation}\label{eq:spinningsupport}
    \forall1\leq p\leq N:\quad\sum_{\substack{q=1\\q\neq p}}^N(\gamma_{pq}-\eta_{pq'})=\Delta_p,\quad\sum_{\substack{q=1\\q\neq p}}^N(\eta_{p'q}+\eta_{p'q'})=J_p.
\end{equation}
Spinning operators also satisfy the ``gauge invariance'' condition \eqref{eq:confgauge}. In Mellin space, this is reflected in the constraint\footnote{When written in terms of $\eta$, the shifts are actually in the opposite direction. For example, $[\mathcal M]_{p'q}=\mathcal M(\gamma_{p'q}\mapsto\gamma_{p'q}-1)\equiv\mathcal M(\eta_{p'q}\mapsto\eta_{p'q}+1)$. }
\begin{equation}\label{eq:gauge}
    \forall1\leq p\leq N:\quad \sum_{\substack{q=1\\q\neq p}}^N\left(\gamma_{pq}[\mathcal M]^{pq}_{p'q}-\eta_{pq'}[\mathcal M]^{pq'}_{p'q'}\right)=0.
\end{equation}
Note that after performing the shifts, the Mellin variables in the resulting function live on a different support, which satisfies \eqref{eq:spinningsupport} with $\Delta_p\mapsto \Delta_p-1$ and $J_p\mapsto J_p-1$\footnote{One way to understand this is to note that the function $X_p\cdot\partial_{Z_p}G(X_1,Z_1,\cdots,X_N,Z_N)$ in \eqref{eq:confgauge} has conformal dimension $\Delta_p-1$ and spin $J_p-1$ with respect to the $p$-th operator.}. We will discuss the support issues in detail in section \ref{subsec:covariant}.

Using the primed notation \eqref{eq:prime1} and \eqref{eq:prime2}, the definition of the spinning Mellin amplitude can be written in a unified way:
\begin{equation}\label{eq:2nscalar}
    G(X_1,X_{1'},\cdots,X_N,X_{N'})=\sum_{\eta\in\mathbb Z}\int[{\rm d}\gamma]\,\mathcal M(\gamma,\eta)\prod_{\substack{r,s\in\mathcal S\\r\prec s}}\begin{cases}
        \displaystyle\frac{\Gamma(\gamma_{rs})}{(-X_{rs})^{\gamma_{rs}}},&\text{$r,s$ unprimed},\\\\
        \displaystyle\frac{(X_{rs})^{\eta_{rs}}}{\eta_{rs}!},&\text{otherwise},
    \end{cases}
\end{equation}
where $\mathcal S=\{1,1',2,2',\cdots,N,N'\}$ is the set of all external labels, and $r\prec s$ means comparing $r$ and $s$ after dropping all primes (e.g., $1'\prec2$). The constraints on the Mellin variables now combine into one:
\begin{equation}\label{eq:mcFinal}
    \forall r\in\mathcal S:\quad\sum_{\substack{s\in\mathcal S\\s\neq r}}\gamma_{rs}=\Delta_r,
\end{equation}
and can be written in terms of auxiliary momenta as 
\begin{equation}
    \gamma_{rs}=\mathtt k_r\cdot\mathtt k_s,\quad \sum_{s\in \mathcal S}\mathtt k_s=0,\quad\mathtt k_s^2=-\Delta_s,\quad \mathtt k_p\cdot \mathtt k_{p'}=0.
\end{equation}
The gauge invariance constraint becomes
\begin{equation}\label{eq:gaugeFinal}
    \forall1\leq p\leq N:\quad \sum_{\substack{s\in\mathcal S\\s\neq p,p'}}\gamma_{ps}[\mathcal M]^{ps}_{p's}=0.
\end{equation}
One should keep in mind that the unprimed variables $\gamma_{pq},\Delta_p$ are continuous, while the primed variables $\gamma_{p'q},\gamma_{p'q'},\Delta_{p'}$ are discrete and take only integer values.

Definition \eqref{eq:2nscalar} suggests treating the $N$-point spinning Mellin amplitude as a $2N$-point scalar Mellin amplitude, where the spinning operator $\mathcal O(X_p,Z_p)= \mathcal O(X_p,X_{p'})$ can be viewed as two fictitious scalars $\mathcal O(X_p)$ and $\mathcal O(X_{p'})$ at null-separated positions $X_p\cdot X_{p'}=0$. Such a statement can be made rather precise when considering the factorization of spinning Mellin amplitudes. Consider the Casimir equation for the spinning amplitude in the OPE channel
\begin{equation}
    \langle\mathcal O(X_1,X_{1'})\cdots\mathcal O(X_\ell,X_{\ell'})|\mathcal O_{\Delta,J}|\mathcal O(X_{\ell+1},X_{(\ell+1)'})\cdots\mathcal O(X_N,X_{N'})\rangle.
\end{equation}
The corresponding Casimir operator now acts on both the position $X$ and the polarization $Z$ on the left half:
\begin{equation}
    \begin{aligned}
        \mathcal C =&\ \frac12\left[\sum_{a=1}^\ell \left(X_{aM}\frac{\partial}{\partial X_a^N}-X_{aN}\frac{\partial}{\partial X_a^M} + Z_{aM}\frac{\partial}{\partial Z_A^N}-Z_{aN}\frac{\partial}{\partial Z_A^M}\right)\right]^2 \\
        =&\ \frac12\left[\sum_{a\in \mathcal L}\left(X_{aM}\frac{\partial}{\partial X_a^N}-X_{aN}\frac{\partial}{\partial X_a^M}\right)\right]^2,
    \end{aligned}
\end{equation}
where $\mathcal L=\{1,1',\cdots,\ell,\ell'\}$ contains all external labels on the left (and similarly for $\mathcal R$). This Casimir operator acts in the same way on both $X_a$ and $X_{a'}$. As a result, through the relation \eqref{eq:similar} between the Mellin kernels, the Casimir operator in Mellin space takes the same form as in scalar amplitudes \eqref{eq:casMelscalar}:
\begin{equation}\label{eq:casMel}
    \hat{\mathbb C}\mathcal M=\gamma_{LR}(d-\gamma_{LR})\mathcal M+\sum_{\substack{a,b\\a\neq b}}\sum_{\substack{i,j\\i\neq j}}\left(\gamma_{ai}\gamma_{bj}(\mathcal M-[\mathcal M]_{aj,bi}^{ai,bj})+\gamma_{ab}\gamma_{ij}[\mathcal M]_{ai,bj}^{ab,ij}\right),
\end{equation}
where the only difference is that $a,b\in \mathcal L$ and $i,j\in \mathcal R$ now range over both unprimed and primed labels. Since the equation is identical to the Casimir equation for $2N$-point scalar Mellin amplitudes, we can directly borrow the factorization formula from the scalar case. 
For instance, for the exchange of a scalar operator with dimension $\Delta$, the spinning Mellin amplitudes factorize as
\begin{equation}\label{eq:spinpole}
    \mathcal M(\gamma,\eta)\sim\frac{\mathcal Q_m}{\gamma_{LR}-\tau-2m},\quad m=0,1,2,\cdots,\quad
\end{equation}
where the residues are still given by
\begin{equation}\label{eq:scalarfacpreview}
    \mathcal Q_m = \frac{-2\Gamma(\Delta)m!}{(1+\Delta-\frac{d}{2})_m}\left(\frac{1}{2^m m!}  \hat{\mathbbm x}^m  \circ \mathcal M_{\mathcal L}\right) \left(\frac{1}{2^m m!}  \hat{\mathbbm x}^m  \circ \mathcal M_{\mathcal R}\right),
\end{equation}
and the ``Mandelstam'' variable is
\begin{equation}
    \gamma_{LR}=\sum_{a,i}\gamma_{ai}=\sum_{b=1}^\ell \sum_{j=\ell+1}^N \left(\gamma_{bj}-\eta_{b'j}-\eta_{bj'}-\eta_{b'j'}\right).
\end{equation}
This is indeed the correct factorization formula for the $N$-point spinning Mellin amplitudes defined in \eqref{eq:spinMellin}!

We will justify the pole structure \eqref{eq:spinpole} in section~\ref{sec:opepole} through a careful OPE analysis. The correctness of our spinning factorization formula is also supported by concrete examples in section~\ref{sec:boots}.

Finally, we remark that there is nothing special about symmetric traceless operators in our definition. For general mixed-symmetry correlators, one simply Mellin transforms all scalar products, and a spinning operator with $h$ polarizations in the correlator can be viewed as $h+1$ scalars. For example, for spin-$(J_{(1)},J_{(2)})$ operators, there are two polarizations
\begin{equation}
    X_{p'}:=Z_{p,(1)},\quad X_{p''}:=Z_{p,(2)}.
\end{equation}
We define the corresponding Mellin amplitude in a similar way as
\begin{equation}
    \begin{aligned}
        &G(X_1,X_{1'},X_{1''},X_2,X_{2'},X_{2''},\cdots,X_\ell,X_{\ell'},X_{\ell''})\\
        =\,&\sum_{\eta\in\mathbb Z}\int[{\rm d}\gamma]\,\mathcal M(\gamma,\eta)\prod_{\substack{r,s\in\mathcal S\\r\prec s}}\begin{cases}
        \displaystyle\frac{\Gamma(\gamma_{rs})}{(-X_{rs})^{\gamma_{rs}}},&\text{$r,s$ unprimed},\\\\
        \displaystyle\frac{(X_{rs})^{\eta_{rs}}}{\eta_{rs}!},&\text{otherwise},
    \end{cases}
    \end{aligned}
\end{equation}
where $\mathcal S = \{1,1',1'',2,2',2'',\cdots,\ell,\ell',\ell''\}$ contains all external labels. The Mellin variables are subject to 
\begin{equation}
    \forall r\in\mathcal S:\quad\sum_{\substack{s\in\mathcal S\\s\neq r}}\gamma_{rs}=\Delta_r,
\end{equation}
and the amplitudes satisfy a series of gauge invariance constraints related to \eqref{eq:mixgauge}:
\begin{equation}
    \forall1\leq p\leq N:\quad \sum_{\substack{s\in\mathcal S\\s\neq p,p',p''}}\gamma_{ps}[\mathcal M]^{ps}_{p's}=\sum_{\substack{s\in\mathcal S\\s\neq p,p',p''}}\gamma_{ps}[\mathcal M]^{ps}_{p''s}=\sum_{\substack{s\in\mathcal S\\s\neq p,p',p''}}\gamma_{p's}[\mathcal M]^{p's}_{p''s}=0.
\end{equation}
The factorization of these Mellin amplitudes then resembles that of $3N$-point scalar amplitudes.

\subsection{The relation to single-spin Mellin amplitudes}\label{subsec:singlespin}

In principle, we can also adapt the factorization formulas for spin-$J$ ($J=1,2$) operators in~\cite{Goncalves:2014rfa} to our spinning Mellin amplitudes. These formulas involve single-spin Mellin amplitudes (the only spinning operator is the exchanged operator). However, our definition of spinning amplitudes, when restricted to single-spin ones, differs from that in~\cite{Goncalves:2014rfa} both in its Mellin kernel and in the operator normalization~\eqref{eq:twoPointNormalization}. Although in section~\ref{sec:factorization} we will directly derive the factorization formula within our framework, it is still useful to compare these two definitions of single-spin amplitudes.

Consider the single-spin Mellin amplitude in our definition, which will later be identified with the left half-amplitude:
\begin{equation}\label{eq:halfLdef}
    \begin{aligned}
        &G_{\mathcal L}(X_1,X_2,\cdots,X_\ell,X_L,Z_L)\\
        =\,&\sum_{\eta\in\mathbb Z}\int[{\rm d}\gamma]\,\mathcal M_{\mathcal L}(\gamma,\eta)\prod_{\substack{a,b\\a<b}}\frac{\Gamma(\gamma_{ab})}{(-X_{ab})^{\gamma_{ab}}}\prod_a\frac{\Gamma(\gamma_{aL})}{(-X_{aL})^{\gamma_{aL}}}\prod_a\frac{(2X_a\cdot Z_L)^{\eta_{aL'}}}{\eta_{aL'}!}.
    \end{aligned}
\end{equation}
As usual, the range of $a,b$ is over all external labels on the left: $\mathcal L=\{1,\cdots,\ell\}$. Suppose the exchanged operator $\mathcal O(X_L,Z_L)$ has dimension $\Delta$ and spin $J$. The Mellin variables satisfy momentum conservation:
\begin{equation}\label{eq:internalSupport}
    \sum_a\gamma_{aL}=\Delta,\quad\sum_a\eta_{aL'}=J,\quad\quad\forall1\leq a\leq\ell:\ \gamma_{aL}-\eta_{aL'}+\sum_{\substack{b\\b\neq a}}\gamma_{ab}=\Delta_a,
\end{equation}
and the half-amplitude $\mathcal M_{\mathcal L}(\gamma,\eta)$ satisfies the gauge invariance condition
\begin{equation}
    \sum_a\gamma_{aL}[\mathcal M_{\mathcal L}]_{aL'}^{aL}=0.\label{eq:singleGauge}
\end{equation}

To relate our definition to that in \cite{Goncalves:2014rfa}, we note that the correlator with the spinning operator $\mathcal O_{\text{\cite{Goncalves:2014rfa}}}$ and the same scalar operators is $2^{-J/2}G_{\mathcal L}$. Their single-spin Mellin amplitude $\widetilde{\mathcal M}_{\mathcal L}^{c_1\cdots c_J}$ has multiple components and is defined for each polarization structure as
\begin{equation}\label{eq:1410def}
    \begin{aligned}
        &2^{-J/2}G_{\mathcal L}(X_1,\cdots,X_\ell,X_{L},Z_L)\\
        =\,&\sum_{c_1,\cdots,c_J\in\mathcal L}(Z_L\cdot X_{c_1})(Z_L\cdot X_{c_2})\cdots (Z_L\cdot X_{c_J}) \\
        &\qquad \qquad  \times \int[{\rm d}\tilde\gamma]\widetilde{\mathcal M}_{\mathcal L}^{c_1\cdots c_J}(\tilde\gamma) \prod_{\substack{a,b\\a<b}}\frac{\Gamma(\tilde\gamma_{ab})}{(-X_{ab})^{\tilde\gamma_{ab}}}\prod_a\frac{\Gamma(\tilde\gamma_{aL}+n_a)}{(-X_{aL})^{\tilde\gamma_{aL}+n_a}},
    \end{aligned}
\end{equation}
where $n_a$ denotes the number of occurrences of $a$ in the sequence $\{c_1,\cdots,c_J\}$. For this definition, the momentum conservation reads
\begin{equation}
    \sum_a\tilde\gamma_{aL}=\Delta-J=\tau,\quad\quad\forall1\leq a\leq\ell:\ \tilde\gamma_{aL}+\sum_{\substack{b\\b\neq a}}\tilde\gamma_{ab}=\Delta_a,
\end{equation}
and the gauge invariance condition is 
\begin{equation}\label{eq:gauge1410}
    \sum_a (\tilde\gamma_{aL}+n_a)\mathcal {\widetilde M}_{\mathcal L}^{ac_2\cdots c_J}=0.
\end{equation}
By construction, $\widetilde{\mathcal M}_{\mathcal L}^{c_1\cdots c_J}$ is symmetric under permuting $\{c_1,\cdots,c_J\}$. Therefore, we can further collect terms in \eqref{eq:1410def} that have the same power of $Z_L\cdot X$. For a product with given powers $(Z_L\cdot X_1)^{n_1}(Z_L\cdot X_2)^{n_2}\cdots (Z_L\cdot X_\ell)^{n_\ell}$, there are $\frac{J!}{n_1!\cdots n_\ell!}$ such terms in the sum, and we can rewrite \eqref{eq:1410def} as
\begin{equation}
    \begin{aligned}
        &2^{-J/2}G_{\mathcal L}(X_1,\cdots,X_\ell,X_{L},Z_L)\\
        =\,&\sum_{\substack{n_1,\cdots,n_\ell \geq 0\\ n_1+\cdots+n_\ell=J}} 2^{-J}\frac{J!}{n_1!\cdots n_\ell!}(2Z_L\cdot X_1)^{n_1}(2Z_L\cdot X_2)^{n_2}\cdots (2Z_L\cdot X_\ell)^{n_\ell} \\
        &\qquad \qquad  \times \int[{\rm d}\tilde\gamma]\widetilde{\mathcal M}_{\mathcal L}^{(n_1\cdots n_\ell)}(\tilde\gamma) \prod_{\substack{a,b\\a<b}}\frac{\Gamma(\tilde\gamma_{ab})}{(-X_{ab})^{\tilde\gamma_{ab}}}\prod_a\frac{\Gamma(\tilde\gamma_{aL}+n_a)}{(-X_{aL})^{\tilde\gamma_{aL}+n_a}},
    \end{aligned}
\end{equation}
where 
\begin{equation}
    \widetilde{\mathcal M}_{\mathcal L}^{(n_1\cdots n_\ell)} := \widetilde{\mathcal M}_{\mathcal L}^{\overbrace{\scriptstyle1\cdots1}^{n_1}\cdots\overbrace{\scriptstyle\ell\cdots\ell}^{n_\ell}}.
\end{equation}
We have also rewritten $Z_L\cdot X$ in terms of $2Z_L\cdot X$. Comparing with our definition \eqref{eq:halfLdef} and accounting for the factor $2^{-J/2}$ on the left-hand side, we can identify $\eta_{aL'}=n_a$ and $\gamma_{aL}=\tilde\gamma_{aL}+n_a$ on both sides, which leads to
\begin{equation}\label{eq:values}
    \widetilde{\mathcal M}_{\mathcal L}^{(n_1\cdots n_\ell)}(\tilde\gamma_{ab},\tilde\gamma_{aL}) = \frac{2^{J/2}}{J!}\mathcal M_{\mathcal L}(\gamma_{ab}\mapsto\tilde\gamma_{ab},\gamma_{aL}\mapsto\tilde\gamma_{aL}+n_a,\eta_{aL'}\mapsto n_a).
\end{equation}
Using the shift operators, we can also write\footnote{Actually, by momentum conservation, we can only take $\eta_{aL'}=0$ after performing the shifts. }
\begin{equation}\label{eq:translate1}
    \widetilde{\mathcal M}_{\mathcal L}^{c_1\cdots c_J}(\tilde\gamma_{ab},\tilde\gamma_{aL})= \frac{2^{J/2}}{J!} [{\mathcal M}_{\mathcal L}]_{c_1L',\cdots,c_JL'}^{c_1L,\cdots,c_JL}\Big|_{\eta_{aL'}\mapsto 0,\gamma_{ab}\mapsto\tilde\gamma_{ab},\gamma_{aL}\mapsto\tilde\gamma_{aL}}.
\end{equation}
One can also see that the gauge invariance condition \eqref{eq:gauge1410} is satisfied. 

To represent our single-spin amplitude $\mathcal M_{\mathcal L}$ in terms of $\widetilde{\mathcal M}_{\mathcal L}$, we need to interpolate a function on the lattice points of $\eta$, which takes the value $2^{-J/2}J!\widetilde{\mathcal M}_{\mathcal L}^{(n_1\cdots n_\ell)}$ at the point $(\eta_{1L'},\cdots,\eta_{\ell L'}) = (n_1,\cdots,n_\ell)$. In principle, there are infinitely many such interpolating functions, as discussed in section \ref{subsec:discrete}. Keeping the denominators in the form \eqref{eq:spinpole}, we require each numerator to be homogeneous of factorial degree $J$ (counting $(\eta)^{\underline n}$ as degree $n$). This fixes the numerator polynomials uniquely and gives
\begin{equation}\label{eq:translate2}
    {\mathcal M}_{\mathcal L}(\gamma_{ab},\gamma_{aL},\eta_{aL'})=\frac{J!}{2^{J/2}}\sum_{\substack{n_1,\cdots,n_\ell \geq 0\\ n_1+\cdots+n_\ell=J}}\left(\prod_a\frac{(\eta_{aL'})^{\underline{n_a}}}{n_a!}\right)\widetilde{\mathcal M}_{\mathcal L}^{(n_1\cdots n_\ell)}\Big|_{\tilde\gamma_{ab}\mapsto\gamma_{ab},\tilde\gamma_{aL}\mapsto\gamma_{aL}-\tilde n_a},
\end{equation}
where the factorial power is defined as 
\begin{equation}
    (\eta)^{\underline{n}} := \eta(\eta-1)\cdots(\eta-n+1),
\end{equation}
and $\tilde n_a$ stands for $\eta_{aL'}$ when the original $\tilde\gamma_{aL}$ appears in the denominators, or $n_a$ when it appears in the numerators. The factor $(\eta_{aL'})^{\underline{n_a}}$ in the formula can be thought of as related to $(2Z_L \cdot X_a)^{n_a}$ in position space. In appendix~\ref{app:fac}, we provide a constructive procedure for putting a rational Mellin amplitude, with the prescribed OPE denominators held fixed, into this canonical form.

As an example, consider the correlator $\langle \mathcal O\mathcal O\mathcal O\mathcal J \rangle$ in the holographic model of supergluons in $\rm AdS_5\times S^3$, where $\mathcal O$ is the scalar supergluon operator and $\mathcal J$ is the spin-1 gluon operator (a brief introduction to this model can be found in subsection \ref{subsec:supergluons}). The color-ordered Mellin amplitude of this correlator, up to the overall coefficients and R-symmetry structures, is given in \cite{Cao:2024bky} by
\begin{align}
    \widetilde{\mathcal{M}}_{\mathcal L}^{1}=&\ \widetilde{\mathcal{M}}_{\mathcal L}^{(100)} = \frac{1}{\tilde\gamma_{12}-1} + \frac{1}{\tilde\gamma_{1L}-1},\\
    \widetilde{\mathcal{M}}_{\mathcal L}^{2}=&\ \widetilde{\mathcal{M}}_{\mathcal L}^{(010)} =\frac{1}{\tilde\gamma_{12}-1} - \frac{1}{\tilde\gamma_{1L}-1},\\
    \widetilde{\mathcal{M}}_{\mathcal L}^{3}=&\ \widetilde{\mathcal{M}}_{\mathcal L}^{(001)} =-\frac{1}{\tilde\gamma_{12}-1} - \frac{1}{\tilde\gamma_{1L}-1}.
\end{align}
Following \eqref{eq:translate2}, we simply replace $\tilde\gamma_{12}\mapsto\gamma_{12}$, and replace $\tilde\gamma_{1L}\mapsto\gamma_{1L}-\eta_{1L'}$ since all of them are in the denominators. These amplitudes are then added up together with the prefactor of $\eta$. The result is 
\begin{equation}
    {\mathcal M}_{\mathcal L}(\gamma_{ab},\gamma_{aL},\eta_{aL'})= \frac{1}{\sqrt2}\frac{\eta_{1L'}+\eta_{2L'}-\eta_{3L'}}{\gamma_{12}-1}+\frac{1}{\sqrt2}\frac{\eta_{1L'}-\eta_{2L'}-\eta_{3L'}}{\gamma_{1L}-\eta_{1L'}-1},
\end{equation}
which has the correct s- and t-channel pole structures. One can also check that this amplitude satisfies the gauge invariance condition, and correctly reproduces the corresponding $\widetilde{\mathcal{M}}_{\mathcal L}$ when we put it into \eqref{eq:translate1}.

\section{Pole structures and residues}\label{sec:opepole}

When we separate the external operators of a conformal correlator into two groups and let them approach each other, the correlator factorizes into a product for each exchanged operator in the OPE. This factorization is captured by the pole structures in Mellin amplitudes through the Mellin transform, and the residues on the poles are factorizable accordingly. In this section, we first show how scaling behaviors in position space are related to poles in Mellin space. We then analyze the scaling behaviors that arise in the OPE of generic spinning correlators, which map to corresponding pole structures in spinning Mellin amplitudes. These pole structures and their factorization properties serve as the basis for the derivation of the factorization formula in section \ref{sec:factorization}. The main argument in this section is adapted and refined from \cite{Fitzpatrick:2011ia}.

\subsection{Scales, poles, and pinches}\label{subsec:pinch}

A feature of the (inverse) Mellin transform is that the scales in position space are encoded in the poles of the Mellin amplitudes. This can be illustrated by a toy integral 
\begin{equation}
    \mathcal I_1=\int_{a-i\infty}^{a+i\infty}\frac{{\rm d}\gamma}{2\pi i}f(\gamma)\frac{\Gamma(\gamma)}{x^{\gamma}},\quad a>0.
\end{equation}
Suppose $f(\gamma)$ has no poles and decays sufficiently rapidly at infinity. By the residue theorem, we can close the contour to the left, and the integral picks up poles in the $\Gamma$ function. The result evaluates to 
\begin{equation}
    \mathcal I_1=\sum_{m=0}^\infty\mathop{\rm Res}_{\gamma=-m}\left[f(\gamma)\frac{\Gamma(\gamma)}{x^{\gamma}}\right]=\sum_{m=0}^\infty \frac{(-)^m}{m!}f(-m)x^m,
\end{equation}
from which we see that each pole in the integrand corresponds to a term $x^m$ in the result. Conversely, when there are terms like $x^{\tau+n}$ appearing in the result, we can deduce that $f(\gamma)$ cannot be regular and must contain poles at $\gamma=-\tau-n$. In this case, one can see that the pole structures and the scaling behaviors are directly related. 

For multi-variate Mellin transforms, poles may not be explicitly present in the integrand but can arise from the pinching of other poles. This can be shown by another toy integral
\begin{equation}\label{eq:toyintegral}
    \mathcal I_2=\int_{a-i\infty}^{a+i\infty}\frac{{\rm d}\gamma}{2\pi i} \frac{\Gamma(\gamma_1+\gamma)\Gamma(\gamma_2-\gamma)}{x^{\gamma}},\quad -\gamma_1 < a < \gamma_2,
\end{equation}
which evaluates to 
\begin{equation}
    \mathcal I_2= \sum_{m=0}^\infty\mathop{\rm Res}_{\gamma=-\gamma_1-m}\left[\frac{\Gamma(\gamma_1+\gamma)\Gamma(\gamma_2-\gamma)}{x^{\gamma}}\right]=\sum_{m=0}^\infty \frac{(-)^m}{m!}x^{\gamma_1+m} \Gamma(\gamma_1+\gamma_2+m).
\end{equation}
As a function of $\gamma_1$ and $\gamma_2$, the result has poles located at $\gamma_1+\gamma_2\in\mathbb Z_{\leq0}$, with the same pole structure as $\Gamma(\gamma_1+\gamma_2)$. If $\mathcal I_2$ appears as a sub-integral within a multi-variate Mellin transform of $\gamma_1$ or $\gamma_2$, say 
\begin{equation}
    \mathcal J=\int_{a'-i\infty}^{a'+i\infty}\frac{{\rm d}\gamma_1}{2\pi i} \frac{\Gamma(\gamma_1)}{y^{\gamma_1}}\int_{a-i\infty}^{a+i\infty}\frac{{\rm d}\gamma}{2\pi i} \frac{\Gamma(\gamma_1+\gamma)\Gamma(\gamma_2-\gamma)}{x^{\gamma}},
\end{equation}
then even if the integrand does not explicitly contain a factor of $\Gamma(\gamma_1+\gamma_2)$, effectively there exist ``poles'' located at $\gamma_1=-\gamma_2-m$ for $m=0,1,\dots$. One can see this even before performing the integral: as we take $\gamma_1+\gamma_2\to-m$ in the integrand of \eqref{eq:toyintegral}, the poles of $\Gamma(\gamma_1+\gamma)\Gamma(\gamma_2-\gamma)$ pinch the contour of $\gamma$, giving rise to poles in the result. We denote this pole-pinching mechanism by
\begin{equation}
    \Gamma(\gamma_1+\gamma)\Gamma(\gamma_2-\gamma)\leadsto\Gamma(\gamma_1+\gamma_2).
\end{equation}
Generalizing this observation enables us to recursively analyze the scaling dependence of multi-variate Mellin transforms by identifying all possible ways of pinching explicit poles to produce effective ``poles''. See appendix C of~\cite{Yuan:2018qva} for a systematic discussion with more pedagogical examples and efficient algorithms.

\subsection{Two-particle OPE}\label{sec:ope2}

For simplicity, we will mainly use the usual $d$-dimensional language instead of the embedding formalism to discuss the OPE. 

Let us first consider the OPE of two traceless mixed-symmetry tensors $\mathcal A_{\alpha_1\cdots\alpha_{J_A}}$ and $\mathcal B_{\beta_1\cdots\beta_{J_B}}$, where $\alpha,\beta$ denote\footnote{For simplicity, we use $\alpha$ to denote a generic member of $\alpha_1,\cdots,\alpha_{J_A}$. Similarly, $g_{\alpha\alpha}$ represents an arbitrary metric with both indices in $\alpha$, say $g_{\alpha_1\alpha_2},g_{\alpha_1\alpha_3},\cdots$ collectively.} the Lorentz indices and $J_A,J_B$ denote the ranks (namely, the sum of all spins). As both operators are bosonic, only bosonic operators (traceless mixed-symmetry tensors) appear in the OPE:
\begin{equation}
    \mathcal A_{\alpha_1\cdots\alpha_{J_A}}(x)\mathcal B_{\beta_1\cdots\beta_{J_B}}(0)=\sum_{\mathcal C}F_{\alpha_1\cdots\alpha_{J_A}\beta_1\cdots\beta_{J_B}}^{\gamma_1\cdots\gamma_{J_C}}(x,\partial)\mathcal C_{\gamma_1\cdots\gamma_{J_C}}(0).
\end{equation}
Here $\partial$ generates the descendant operators when acting on $\mathcal C$. We first focus on the primary part of the OPE coefficient $\widetilde{F}(x):=F(x,0)$, which is a tensor that can only depend on the metric and $x$. 

What is the possible structure of $\widetilde{F}$? Due to the traceless condition of $\mathcal C$, the metric $g^{\gamma\gamma}$ cannot appear. Moreover, in index-free notation, the indices ${}_\alpha,{}_\beta$ are dotted into $(z_A)^\alpha,(z_B)^\beta$ respectively\footnote{Here $(z_A)^\alpha$ denotes arbitrary polarization vectors $(z_{A,(i)})^\alpha$ of the mixed-symmetry operator $\mathcal A$. Similarly for $(z_B)^\beta$.}. Since $z_A \cdot z_A = z_B \cdot z_B = 0$, the metrics $g_{\alpha\alpha},\, g_{\beta\beta}$ in $\widetilde{F}$ do not affect the final result and can be safely dropped. Therefore, $\widetilde{F}$ can only depend on 
\begin{equation*}
    \{x^2,x_\alpha,x_\beta,x^\gamma,g_{\alpha\beta},\delta_\alpha^{\ \gamma},\delta_\beta^{\ \gamma}\},
\end{equation*}
from which we can write down 
\begin{equation}\label{eq:Fprimary}
    \widetilde{F}\propto(x^2)^{-k}(x_\alpha)^{m_1}(x_\beta)^{m_2}(x^\gamma)^{m_3}(g_{\alpha\beta})^{n_1}(\delta_\alpha^{\ \gamma})^{n_2}(\delta_\beta^{\ \gamma})^{n_3},
\end{equation}
with the powers to be determined later. Now, for simplicity, we restrict to the case where $\mathcal A,\mathcal B,\mathcal C$ are all symmetric operators (the generalization to the mixed-symmetry case is straightforward). When we perform the OPE in the correlator, in the Poincar\'e section \eqref{eq:poincare} it reads $\langle\mathcal A(x,z_A)\mathcal B(0,z_B)\cdots\rangle$. Dotted with the polarizations, the tensor structures in $\widetilde{F}$ map to scalar products in the discrete Mellin transform:
\begin{equation}
    x_\alpha\leftrightarrow z_A\cdot x\propto(Z_A\cdot X_B),\quad x_\beta\leftrightarrow z_B\cdot x\propto(Z_B\cdot X_A),\quad g_{\alpha\beta}\leftrightarrow z_A\cdot z_B\propto(Z_A\cdot Z_B).
\end{equation}
Therefore we can identify 
\begin{equation}\label{eq:OPEeta}
    m_1=\eta_{A'B},\quad m_2=\eta_{AB'},\quad n_1=\eta_{A'B'}.
\end{equation}
Furthermore, counting spins and dimensions on both sides of the OPE, we have
\begin{subequations}\label{eq:OPEcounting}
    \begin{gather}
        m_1+n_1+n_2=J_A,\quad m_2+n_1+n_3=J_B,\quad m_3+n_2+n_3=J_C,\\
        \Delta_A+\Delta_B=2k-m_1-m_2-m_3+\Delta_C.
    \end{gather}
\end{subequations}
From these constraints we can determine all the powers in \eqref{eq:Fprimary}. In particular,
\begin{equation}
    k=\frac{\tau_A+\tau_B-\tau_C}{2}+\eta_{A'B}+\eta_{AB'}+\eta_{A'B'},
\end{equation}
where $\tau:=\Delta-J$ is the conformal twist of a bosonic local operator (for mixed-symmetry operators the twist is defined by the dimension minus the rank $\Delta-J_{(1)}-\cdots-J_{(h)}$). The scaling behavior of $x^2 \propto (-X_A\cdot X_B)$ corresponds to a pole at $\gamma_{AB}=k$ in the continuous Mellin transform. When $\tau_C-\tau_A-\tau_B\notin 2 \mathbb Z_{\geq0}$, this pole cannot come from $\Gamma$ functions or the pinching mechanism, and must be present in the Mellin amplitude $\mathcal M$. In such a case
\begin{equation}
    \mathcal M(\gamma,\eta) \sim \frac{\text{res.}}{\gamma_{AB}-k} \propto \frac{\text{res.}}{\gamma_{LR}-\tau_C},
\end{equation}
where 
\begin{equation}
    \gamma_{LR} = -(\mathtt k_A+\mathtt k_{A'}+\mathtt k_B+\mathtt k_{B'})^2 = \tau_A+\tau_B-2(\gamma_{AB}-\eta_{A'B}-\eta_{AB'}-\eta_{A'B'})
\end{equation}
is the Mandelstam in the OPE channel we are considering. 

When considering the descendants of $\mathcal C$, we can expand $F$ as a series of $\partial$. For the coefficients of each $\partial$, the counting in \eqref{eq:OPEeta} and \eqref{eq:OPEcounting} still applies, but $\Delta_C$ and $J_C$ should be replaced by the dimension and spin of the descendant operators. Schematically, these descendants are obtained by acting with $\partial_\gamma$ (which takes $\Delta\mapsto\Delta+1$, $J\mapsto J+1$) and then contracting with $g^{\gamma\gamma}$ (which takes $\Delta\mapsto\Delta$, $J\mapsto J-2$), resulting in $\tau=\tau_C+2m$ ($m=0,1,2,\cdots$). Therefore, they contribute to poles at $\gamma_{LR}=\tau_C+2m$ according to our previous discussion. In summary, for the whole conformal multiplet of $\mathcal C$, there exists a sequence of poles in the Mellin amplitude:
\begin{equation}
    \mathcal M(\gamma,\eta) \sim \frac{1}{\gamma_{LR}-\tau_C-2m},\quad m=0,1,2,\cdots,
\end{equation}
as we expect in \eqref{eq:spinpole}.

In the perturbation around generalized free field theories (including the large central charge limit of holographic CFTs), the OPE of $\mathcal A$ and $\mathcal B$ contains a special sector of composite operators, the double-trace operators. They can be viewed as the ``double-particle states'' composed of $\mathcal A$ and $\mathcal B$, and have twist $\tau=\tau_A+\tau_B+2m$ for $m\in \mathbb Z_{\geq 0}$. These operators are generally constructed by inserting several derivatives in the normal ordering of $\mathcal A$ and $\mathcal B$, which we denote schematically as $[\mathcal A\partial\cdots\partial\mathcal B]$. Since $\tau-\tau_A-\tau_B\in 2 \mathbb Z_{\geq0}$, the poles of double-trace operators might not be present in the Mellin amplitudes, but come from the $\Gamma$ functions. Actually, by definition, the three-point function
\begin{equation}
    \langle\mathcal A(x)\mathcal B(0)[\mathcal A \partial\cdots\partial \mathcal B](y)\rangle \sim \partial\cdots\partial \langle\mathcal A(x)\mathcal A(y)\rangle \partial\cdots\partial \langle\mathcal B(0)\mathcal B(y)\rangle
\end{equation}
can be computed by taking derivatives of the product of two-point functions $\langle\mathcal A(x)\mathcal A(y)\rangle$ and $\langle\mathcal B(0)\mathcal B(y)\rangle$. This three-point function is completely regular when we take the OPE limit $x\to 0$. As a result, one can show that the corresponding OPE coefficient only contains $x^2$ with non-negative integer powers, which are precisely captured by the explicit poles of $\Gamma(\gamma_{AB})$ in the Mellin transform. 

\subsection{Multi-particle OPE}

The above discussion readily generalizes to multi-operator OPEs. For example, the three-particle OPE reads, schematically, 
\begin{equation}
     \mathcal A_\alpha(x_1)\mathcal B_\beta(x_2)\mathcal C_\gamma(x_3)=\sum_{\mathcal D}\frac{(g_{\alpha\beta})^{\eta_{1'2'}}(x_{12}^\alpha)^{\eta_{1'2}}\cdots+O(g_{\alpha\alpha},g_{\beta\beta},g_{\gamma\gamma})}{(x_{12}^2)^{\gamma_{12}}(x_{13}^2)^{\gamma_{13}}(x_{23}^2)^{\gamma_{23}}}\mathcal D_\delta(0).
\end{equation}
Generically, such scaling behaviors can only arise from explicit poles in $\mathcal M(\gamma,\eta)$ of the form
\begin{equation}
    \mathcal M(\gamma,\eta)\sim\frac{\text{res.}}{\gamma_{LR}-\tau_D-2m},
\end{equation}
where $\gamma_{LR} = -\left(\sum_{a\in \mathcal L} \mathtt k_a\right)^2$ is the Mandelstam in the corresponding OPE channel, with $\mathcal L=\{1,1',2,2',3,3' \}$ (which may contain indices like $1''$ for mixed-symmetry operators). 

In perturbation theory, single-trace operators correspond to explicit poles of $\mathcal M(\gamma,\eta)$, while multi-trace operators correspond to the effective poles arising from pinching $\Gamma(\gamma_{pq})$ among themselves or with explicit poles in $\mathcal M(\gamma,\eta)$. For example, the poles of $[\mathcal A\mathcal B\mathcal C]$ arise from the pinching 
\begin{equation}
    \Gamma(\gamma_{12})\Gamma(\gamma_{13})\Gamma(\gamma_{23})\leadsto\Gamma(\gamma_{12}+\gamma_{13}+\gamma_{23}),
\end{equation}
while the poles of $[\mathcal E\mathcal C]$ (suppose $\mathcal E$ is a single-trace operator appearing in the OPE of $\mathcal A\times \mathcal B$) arise from pinching\footnote{We remind the reader that $\Gamma(x)$ does not necessarily mean that there is a $\Gamma$ function in the integrand, but simply stands for generic functions with a sequence of poles at $x=0,-1,-2,\cdots$ }
\begin{equation}
    \Gamma\left(\frac{\tau_E-\gamma_{\{AB\}}}{2}\right)\Gamma(\gamma_{13})\Gamma(\gamma_{23})\leadsto\Gamma\left(\frac{\tau_E-\gamma_{\{AB\}}}{2}+\gamma_{13}+\gamma_{23}\right),
\end{equation}
where $\gamma_{\{AB\}}:=\tau_A+\tau_B-2(\gamma_{12}-\eta_{1'2}-\eta_{12'}-\eta_{1'2'})$ is the Mandelstam in the $\mathcal A \times \mathcal B$ OPE, and the sequence of poles $\Gamma\left(\frac{\tau_E-\gamma_{\{AB\}}}{2}\right)$ arises from the Mellin amplitude $\mathcal M$. The pinch leads to the regular behavior $(x_{13}^2)^{-\gamma_{13}}(x_{23}^2)^{-\gamma_{23}}$ with $\gamma_{13},\gamma_{23}\in\mathbb Z_{\leq0}$ as expected, as well as the singular behavior $(x_{12}^2)^{-\gamma_{12}}$ producing $\mathcal E$ with
\begin{equation}
    \gamma_{12}=\frac{\tau_A+\tau_B-\tau_E}2+\eta_{1'2}+\eta_{12'}+\eta_{1'2'}-m.
\end{equation}

\subsection{The Casimir equation}\label{subsec:casimir}

From the previous subsection, we learned that for each single-trace operator $\mathcal O_{\Delta,J}$ with dimension $\Delta$ and rank $J$, the OPE channel
\begin{equation*}
    \langle\mathcal O(X_1,Z_{1})\cdots\mathcal O(X_\ell,Z_{\ell})|\mathcal O_{\Delta,J}|\mathcal O(X_{\ell+1},Z_{\ell+1})\cdots\mathcal O(X_N,Z_{N})\rangle
\end{equation*}
corresponds to a sequence of poles in $\mathcal M(\gamma,\eta)$ of the form
\begin{equation}\label{eq:melPole}
    \mathcal M\sim\frac{\mathcal Q_m}{\gamma_{LR}-\tau-2m},\quad m=0,1,2,\cdots,\quad\gamma_{LR}=\sum_{a}\sum_{i}\gamma_{ai}.
\end{equation}
Recall that we assume the summation ranges of $a,b,\cdots$ and $i,j,\cdots$ are over the sets $\mathcal L$ and $\mathcal R$ respectively, where $\mathcal L$ and $\mathcal R$ consist of all external labels on the left and right, including the primed labels from the polarizations. 

Following \cite{Fitzpatrick:2011ia,Goncalves:2014rfa} and our previous discussions, the residue $\mathcal Q_m$ can be determined by studying the Casimir equation
\begin{equation}
    \hat{\mathbb C} \mathcal M=-c_{\Delta,J}\mathcal M,
\end{equation}
where the action of $\hat{\mathbb C}$ can be found in \eqref{eq:casMel}, and $c_{\Delta,J}$ is the eigenvalue of $\mathcal O_{\Delta,J}$:
\begin{equation}
    c_{\Delta,J} := -\Delta(d-\Delta)+J_{(1)}(d-2+J_{(1)}) + \cdots + J_{(h)}(d-2h+J_{(h)}).
\end{equation}
We mainly focus on the exchange of traceless symmetric operators, for which $c_{\Delta,J}$ reduces to $-\Delta(d-\Delta)+J(d-2+J)$. Evaluating the residue of $\mathcal M$ at $\gamma_{LR}=\Delta-J+2m$ in the Casimir equation, we obtain a recurrence relation for $\mathcal Q_m$:
\begin{equation}\label{eq:continuousCasimir}
    \begin{aligned}
        0=\,&\left(2J(\Delta-1+2m)+2m(d-2\Delta-2m)\right)\mathcal Q_m\\
        +\,&\sum_{\substack{a,b\\a\neq b}}\sum_{\substack{i,j\\i\neq j}}\left(\gamma_{ai}\gamma_{bj}(\mathcal Q_m-[\mathcal Q_m]_{aj,bi}^{ai,bj})+\gamma_{ab}\gamma_{ij}[\mathcal Q_{m-1}]_{aj,bi}^{ab,ij}\right).
    \end{aligned}
\end{equation}

\section{Factorization formulas}\label{sec:factorization}

In this section, we systematically derive the factorization formula for spinning Mellin amplitudes by solving the Casimir recurrence relation \eqref{eq:continuousCasimir}. The techniques employed here largely originate from \cite{Goncalves:2014rfa}, where the authors derived factorization formulas case by case for spin-0,1,2 exchanges in scalar Mellin amplitudes. We are going to solve the same recurrence relation, but the interpretation changes: the indices of the Mellin variables now include polarizations, and the solution yields the factorization formula for spinning amplitudes. The final results are collected in subsection~\ref{subsec:facform}.

We improve upon their technique by identifying a closed algebra of so-called covariant shift operators within the Casimir equation, which will be introduced shortly. These covariant shift operators relieve us of the need to keep track of all the component labels as in \cite{Goncalves:2014rfa}, and also make the complicated combinatorial relations in the recurrence relation more transparent. In the main part of this section, we assume the exchange operator is a symmetric traceless operator $\mathcal O(X,Z)$. The extension to mixed symmetry exchanges will be discussed at the end.

\subsection{Support issues and covariant shifts}\label{subsec:covariant}

We first discuss the important issue of support when dealing with shifts in Mellin amplitudes. As mentioned above, the Mellin amplitude is defined on the support of momentum conservation \eqref{eq:mcFinal}. For example, the following two amplitudes are equivalent:
\begin{equation}
    \mathcal M=0,\qquad \mathcal M'=-\Delta_r+\sum_{\substack{s\in\mathcal S\\r\neq s}}\gamma_{rs}.
\end{equation}
However, when applying the shift operator $[\, \boldsymbol{\cdot}\, ]^{rs}$ which shifts $\gamma_{rs}\mapsto\gamma_{rs}+1$ to these two amplitudes, the results appear to be different:
\begin{equation}
    [\mathcal M]^{rs}=0,\qquad [\mathcal M']^{rs}=-\Delta_r+1+\sum_{\substack{s\in\mathcal S\\r\neq s}}\gamma_{rs}.
\end{equation}
This is because the momentum conservation \eqref{eq:mcFinal} is also shifted by the shift operator, and the supports of Mellin amplitudes are changed. Since $\gamma_{rs}$ appears in the momentum conservation constraints of $\Delta_r$ and $\Delta_s$, the operator $[\, \boldsymbol{\cdot}\, ]^{rs}$ effectively shifts the support to $\Delta_r\mapsto \Delta_r-1$ and $\Delta_s\mapsto \Delta_s-1$ (while $[\, \boldsymbol{\cdot}\, ]_{rs}$ shifts the support to $\Delta_r\mapsto \Delta_r+1$ and $\Delta_s\mapsto \Delta_s+1$). Note that when $r$ or $s$ is a primed index, the spins also get shifted but in the opposite direction since $\Delta_{r'}\equiv -J_{r}$. This explains the support of the gauge invariance constraint \eqref{eq:gauge}, which is on $\Delta_p-1$ and $J_p-1$. 

Since we will use the shift operators very frequently in the derivation, it is crucial to keep track of the support on which an expression is defined. Especially, we use $\mathscr M_{\mathcal L}^{\Delta,J}$ to denote the space of left half-amplitudes such that the exchanged operator $\mathcal O(X_L,Z_L)$ has dimension $\Delta$ and spin $J$. For instance, the half-amplitude itself $\mathcal M_{\mathcal L}$ lives in $\mathscr M_{\mathcal L}^{\Delta,J}$, while the gauge invariance condition \eqref{eq:singleGauge} is an equality in $\mathscr M_{\mathcal L}^{\Delta-1,J-1}$. We also note that the single-spin amplitudes $\widetilde{\mathcal M}_{\mathcal L}$ defined in \cite{Goncalves:2014rfa} are functions in $\mathscr M_{\mathcal L}^{\Delta-J,0}$. 

In factorization formulas, there are usually shift operators acting on the half-amplitudes (e.g., see \eqref{eq:facdemoscalar}). However, simple shift operators like $[\, \boldsymbol{\cdot}\,  ]^{ab}$ may change the supports of Mellin amplitudes and effectively affect the external dimensions and spins ($\Delta_a\mapsto \Delta_a-1$, $\Delta_b\mapsto \Delta_b-1$). In order to keep the external dimensions and spins intact in the factorization formula, we introduce the following covariant shift operators:
\begin{equation}\label{eq:covL}
    \begin{aligned}
        &[\, \boldsymbol{\cdot}\,]^{aL}_{aL'}&:&\ \mathscr M_{\mathcal L}^{\Delta,J} \to \mathscr M_{\mathcal L}^{\Delta-1,J-1},\\
        &[\, \boldsymbol{\cdot}\,]^{ab}_{aL,bL}&:&\ \mathscr M_{\mathcal L}^{\Delta,J} \to \mathscr M_{\mathcal L}^{\Delta+2,J},\\
        &[\, \boldsymbol{\cdot}\,]^{ab}_{aL,bL'}&:&\ \mathscr M_{\mathcal L}^{\Delta,J} \to \mathscr M_{\mathcal L}^{\Delta+1,J-1},\\
        &[\, \boldsymbol{\cdot}\,]^{ab}_{aL',bL'}&:&\ \mathscr M_{\mathcal L}^{\Delta,J} \to \mathscr M_{\mathcal L}^{\Delta,J-2}.
    \end{aligned}
\end{equation}
These operators (and their inverses) always maintain the supports of external operators, at the cost of shifting the effective dimension and spin of the exchanged operator. This is acceptable and welcome in the factorization formula, where only the twist $\tau$ of the exchanged operator appears in the pole structures, and gets shifted to $\tau+2m$ for descendant poles. The covariant shift operators make it much easier to keep track of the support issues. As an example, the operator $\hat{\mathbbm x}$ defined in \eqref{eq:xhat} should be understood as 
\begin{equation}
    \hat{\mathbbm x}:=\sum_{\substack{a,b\\a\neq b}}\gamma_{ab}[\, \boldsymbol{\cdot}\,]^{ab}_{aL,bL}.
\end{equation}
Applying $\hat{\mathbbm x}^m$ to $\mathcal M_{\mathcal L} \in \mathscr M_{\mathcal L}^{\Delta,0}$ indeed brings the result into the function space $\mathscr M_{\mathcal L}^{\tau+2m,0}$, compatible with the pole structure.

\subsection{Function space in the factorization formula}

We expect the factorization formula to take the schematic form:
\begin{equation}\label{eq:facform}
    \begin{aligned}
        \mathcal Q_m(\gamma_{ab},\gamma_{ij},\gamma_{ai})&\sim \text{(shift operators)} \circ \Big(\mathcal M_{\mathcal L}(\gamma_{ab},\gamma_{aL},\eta_{aL'}) \mathcal M_{\mathcal R}(\gamma_{ij},\gamma_{iR},\eta_{iR'})\Big).
    \end{aligned}
\end{equation}
We see that it relates the residue $\mathcal Q_m(\gamma_{ab},\gamma_{ij},\gamma_{ai})$ with the half-amplitudes $\mathcal M_{\mathcal L}(\gamma_{ab},\gamma_{aL},\eta_{aL'})$ and $\mathcal M_{\mathcal R}(\gamma_{ij},\gamma_{iR},\eta_{iR'})$, which depend on different sets of Mellin variables at first glance. In order to formulate the factorization formula properly, we must be able to sensibly compare functions depending on $\gamma_{aL},\eta_{aL'},\gamma_{iR},\eta_{iR'}$ with functions depending on $\gamma_{ai}$. Motivated by the interpretation of Mellin variables as Lorentz products of auxiliary momenta, we impose the following relations between the variables $\gamma_{aL},\gamma_{iR}$ and $\gamma_{ai},\gamma_{LR}$:
\begin{subequations}\label{eq:onshellfinal}
    \begin{gather}
        \forall a:\ \gamma_{aL}=\sum_i\gamma_{ai},\label{eq:gammaaL}\\
        \forall i:\ \gamma_{iR}=\sum_a\gamma_{ai},\label{eq:gammaaR}\\
        \gamma_{LR} = \sum_a\gamma_{aL}=\sum_i\gamma_{iR}=\sum_{a,i}\gamma_{ai}.
    \end{gather}
\end{subequations}
We can now present the product $\mathcal M_{\mathcal L}\mathcal M_{\mathcal R}$ as a function that explicitly involves all the Mellin variables ($\gamma_{ab},\gamma_{aL},\eta_{aL'},\gamma_{ij},\gamma_{iR},\eta_{iR'},\gamma_{ai},\gamma_{LR}$) through substitution of the above relations. However, $\mathcal M_{\mathcal L}\mathcal M_{\mathcal R}$ is not a generic function of these variables satisfying these relations. In particular, the value of $\gamma_{LR}$ is not generic but fixed by the support of $\mathcal M_{\mathcal L/\mathcal R}\in\mathscr M_{\mathcal L/\mathcal R}^{\Delta,J}$ to be $\gamma_{LR}=\Delta$:
\begin{subequations}\label{eq:def}
\begin{align}
    \forall a:\quad&\gamma_{aL}-\eta_{aL'}+\sum_{\substack{b\\b\neq a}}\gamma_{ab}=\Delta_a,\label{eq:onshellL}\\
    &\sum_a\gamma_{aL}=\Delta,\quad\sum_a\eta_{aL'}=J,\label{eq:suppL}\\
    \forall i:\quad&\gamma_{iR}-\eta_{iR'}+\sum_{\substack{j\\j\neq i}}\gamma_{ij}=\Delta_i,\label{eq:onshellR}\\
    &\sum_i\gamma_{iR}=\Delta,\quad\sum_i\eta_{iR'}=J.\label{eq:suppR}
\end{align}
\end{subequations}
This suggests we consider the function space $\mathscr Q^{\Delta,J}$ of functions
\begin{equation*}
    \mathcal Q(\gamma_{ab},\gamma_{aL},\eta_{aL'},\gamma_{ij},\gamma_{iR},\eta_{iR'},\gamma_{ai},\gamma_{LR})
\end{equation*}
on the support of \eqref{eq:onshellfinal} and \eqref{eq:def}. For example, the product $\mathcal M_{\mathcal L}\mathcal M_{\mathcal R}\in\mathscr Q^{\Delta,J}$. Also, we expect that the shift operators will map the product to a function $\widetilde{\mathcal Q}_m\in\mathscr Q^{\tau+2m,\text{something}}$, because $\mathcal Q_m$ is the residue at $\gamma_{LR}=\tau+2m$. We will now show that the ``something'' has to be zero and relate $\widetilde{\mathcal Q}_m\in\mathscr Q^{\tau+2m,0}$ to the residue $\mathcal Q_m$.

First, note that substituting \eqref{eq:gammaaL} into \eqref{eq:onshellL} implies that $\widetilde{\mathcal Q}_m$ satisfies
\begin{equation*}
    \forall a:\quad-\Delta_a+\sum_{\substack{b\\b\neq a}}\gamma_{ab}+\sum_i\gamma_{ai}=\eta_{aL'},
\end{equation*}
which is not quite compatible with the momentum conservation
\begin{equation*}
    \forall a:\quad\mathtt k_a\cdot\left(\sum_a\mathtt k_a+\sum_i\mathtt k_i\right)=0
\end{equation*}
we expect for $\mathcal Q_m$ in the full amplitude. On the other hand, $\widetilde{\mathcal Q}_m$ still depends on the variables $\eta_{aL'},\eta_{iR'}$ since $\mathcal M_{\mathcal L}\mathcal M_{\mathcal R}$ depends on them. This is not the expected dependence for the full amplitude, either, because $\mathcal Q_m$ should depend on $\gamma_{ab},\gamma_{ij},\gamma_{ai}$ only. Both requirements are satisfied by evaluating $\widetilde{\mathcal Q}_m$ on the lattice point where $\eta_{aL'}=0$ and $\eta_{iR'}=0$, which implies that $\widetilde{\mathcal Q}_m\in\mathscr Q^{\tau+2m,0}$. In summary,
\begin{equation}
        \mathcal Q_m=\underbrace{\text{(shift operators)} \circ\Big(\mathcal M_{\mathcal L}\mathcal M_{\mathcal R}\Big)}_{\widetilde{\mathcal Q}_m\in\mathscr Q^{\tau+2m,0}}\Big|_{\substack{\eta_{aL'}=0\\\eta_{iR'}=0}}.
\end{equation}

Similar to the covariant shift operators for the function space $\mathscr M_{\mathcal L/\mathcal R}^{\Delta,J}$, we introduce the following covariant shift operators for $\mathscr Q^{\Delta,J}$:
\begin{equation}\label{eq:shifts}
    \begin{aligned}
        \langle\mathcal Q\rangle^{ai}&:=[\mathcal Q]^{aL\,,\;iR,\;ai,LR}_{aL',\,iR'},&\qquad\langle\, \boldsymbol{\cdot}\, \rangle^{ai}:&\ \mathscr Q^{\Delta,J}\to\mathscr Q^{\Delta-1,J-1},\\
        \langle\mathcal Q\rangle_{ai}&:=[\mathcal Q]_{aL\;,\;iR,\;ai,LR}^{aL',\,iR'},&\langle\, \boldsymbol{\cdot}\, \rangle_{ai}:&\ \mathscr Q^{\Delta,J}\to\mathscr Q^{\Delta+1,J+1},\\
        \{\mathcal Q\}^{ab,ij}&:=[\mathcal Q]^{ab,\ ij}_{aL',bL',\,iR',jR'},&\{\, \boldsymbol{\cdot}\, \}^{ab,ij}:&\ \mathscr Q^{\Delta,J}\to\mathscr Q^{\Delta,J-2},\\
        \{\mathcal Q\}_{ab,ij}&:=[\mathcal Q]_{ab,\; ij}^{aL',bL',\,iR',jR'},&\{\, \boldsymbol{\cdot}\, \}_{ab,ij}:&\ \mathscr Q^{\Delta,J}\to\mathscr Q^{\Delta,J+2}.
    \end{aligned}
\end{equation}
Let us explain our notation, taking the definition of $\langle\mathcal Q\rangle^{ai}$ as an example. The superscripts instruct us to increment all appearances of $\gamma_{aL},\gamma_{iR},\gamma_{ai},\gamma_{LR}$ in a given presentation of $\mathcal Q$, so that the relations \eqref{eq:onshellfinal} are preserved. The subscripts increment $\eta_{aL'}$ and $\eta_{iR'}$, restoring the on-shell condition for the external particle $a$ in \eqref{eq:onshellL}. The net result of these shifts is that the support \eqref{eq:suppL} is effectively shifted as $\Delta\mapsto\Delta-1$ and $J\mapsto J-1$. We can now write the Casimir recurrence relation in terms of covariant shifts:
\begin{equation}\label{eq:covRecur}
    \left(2J(\Delta-1+2m)+2m(d-2\Delta-2m)\right)\mathcal Q_m+\hat{\mathbb F}\mathcal Q_m+\hat{\mathbb X}\mathcal Q_{m-1}=0,
\end{equation}
where 
\begin{equation}
    \begin{aligned}
        \hat{\mathbb F}\mathcal Q&:=\sum_{\substack{a,b\\a\neq b}}\sum_{\substack{i,j\\i\neq j}}\gamma_{ai}\gamma_{bj}(\mathcal Q-\langle\mathcal Q\rangle_{aj,bi}^{ai,bj}),&&\hat{\mathbb F}:\ \mathscr Q^{\Delta,J}\to\mathscr Q^{\Delta,J},\\
        \hat{\mathbb X}\mathcal Q&:=\sum_{\substack{a,b\\a\neq b}}\sum_{\substack{i,j\\i\neq j}}\gamma_{ab}\gamma_{ij}\{\langle\mathcal Q\rangle_{aj,bi}\}^{ab,ij},&&\hat{\mathbb X}:\ \mathscr Q^{\Delta,J}\to\mathscr Q^{\Delta+2,J}.
    \end{aligned}
\end{equation}

Finally, let us discuss the important factorizable subspace $\mathscr M_{\mathcal L}^{\Delta,J}\otimes\mathscr M_{\mathcal R}^{\Delta,J}\subset\mathscr Q^{\Delta,J}$ of functions such as the product $\mathcal M_{\mathcal L}\mathcal M_{\mathcal R}$, which only depend on $\gamma_{ai}$ through the particular combinations $\gamma_{aL},\gamma_{iR}$. For a factorizable function, it is always possible to write it in a factorized form where the variables $\gamma_{ai},\gamma_{LR}$ do not appear. As a result, when restricted to the factorizable subspace, the covariant shifts of $\mathscr Q^{\Delta,J}$ factorize:
\begin{equation}\label{eq:factorizability}
    \langle\mathcal F_{\mathcal L}\otimes\mathcal F_{\mathcal R}\rangle^{ai}=[\mathcal F_{\mathcal L}]^{aL}_{aL'}\otimes[\mathcal F_{\mathcal R}]^{iR}_{iR'},\quad\{\mathcal F_{\mathcal L}\otimes\mathcal F_{\mathcal R}\}^{ab,ij}=[\mathcal F_{\mathcal L}]^{ab}_{aL',bL'}\otimes[\mathcal F_{\mathcal R}]^{ij}_{iR',jR'},
\end{equation}
and similarly for the operators with subscripts.

\subsection{Lessons from scalar amplitude factorization}\label{subsec:lessons}

We briefly review the results and conjectures obtained in \cite{Goncalves:2014rfa}, by simply translating equations in that paper into our definition of spinning Mellin amplitudes using \eqref{eq:translate1} and \eqref{eq:translate2}. This will give us a taste of the general structure of solutions as well as some hints on how to proceed systematically.

\begin{itemize}

    \item {\bf Primary factorization of spin-$J$ exchanges}
    
        In appendix B of \cite{Goncalves:2014rfa}, the authors conjectured a formula for the primary residue $\mathcal Q_0$ that applies to any spin-$J$ exchange:
        \begin{equation}
            \mathcal Q_0=\kappa_{\Delta J}J!\sum_{\substack{n_{ai}\geq0\\\Sigma_{i}n_{ai}=n_a\\ \Sigma_{a}n_{ai}=n_i}}\left(\prod_{a,i}\frac{(\gamma_{ai})_{n_{ai}}}{n_{ai}!}\right)\widetilde{\mathcal M}_{\mathcal L}^{(n_1\cdots n_\ell)}\widetilde{\mathcal M}_{\mathcal R}^{(n_{\ell+1}\cdots n_N)},
        \end{equation}
        where
        \begin{equation}
            \kappa_{\Delta J}=(-2)^{1-J}(\Delta+J-1)\Gamma(\Delta-1).
        \end{equation}
        These $n_{ck}$ actually count the number of occurrences of the pair $(c,k)$ in the sequence $\{(c_1,k_1),\cdots,(c_J,k_J)\}$ appearing in the component amplitudes $\widetilde{\mathcal M}_{\mathcal L}^{c_1\cdots c_J}\widetilde{\mathcal M}_{\mathcal R}^{k_1\cdots k_J}$. Using \eqref{eq:translate1}, it is not hard to show that
        \begin{equation}\label{eq:primaryFac}
            \mathcal Q_0=\frac{2^J\kappa_{\Delta J}}{(J!)^2} \left.\hat{\mathbb N}^J \circ (\mathcal M_{\mathcal L}\mathcal M_{\mathcal R})\right|_{\substack{\eta_{aL'}=0\\\eta_{iR'}=0}},
        \end{equation}
        where
        \begin{equation}
            \hat{\mathbb N}\mathcal Q:=\sum_{a,i}\gamma_{ai}\langle\mathcal Q\rangle^{ai},\quad\hat{\mathbb N}:\ \mathscr Q^{\Delta,J}\to\mathscr Q^{\Delta-1,J-1}.
        \end{equation}
        
    \item {\bf Descendant factorization of scalar exchanges}
    
        In \cite{Goncalves:2014rfa}, the authors showed that for scalar exchanges,
        \begin{equation}
            \mathcal Q_m=\kappa_{\Delta0}\frac{m!}{(1-\frac d2+\Delta)_m}\widetilde{\mathcal M}_{\mathcal L,m}\widetilde{\mathcal M}_{\mathcal R,m},
        \end{equation}
        where (in the non-covariant notation of that paper)
        \begin{equation}
            \widetilde{\mathcal M}_{\mathcal L,m}=\frac1{2m}\sum_{\substack{a,b\\a\neq b}}\gamma_{ab}[\widetilde{\mathcal M}_{\mathcal L,{m-1}}]^{ab},\quad \widetilde{\mathcal M}_{\mathcal R,m}=\frac1{2m}\sum_{\substack{i,j\\i\neq j}}\gamma_{ij}[\widetilde{\mathcal M}_{\mathcal R,{m-1}}]^{ij}.
        \end{equation}
        It is not hard to show that, in terms of covariant shifts, this is equivalent to
        \begin{equation}\label{eq:scalarFac}
            \mathcal Q_m=\frac{\kappa_{\Delta0}}{4^mm!(1-\frac d2+\Delta)_m}\left.\hat{\mathbb X}^m \circ(\mathcal M_{\mathcal L}\mathcal M_{\mathcal R})\right|_{\substack{\eta_{aL'}=0\\\eta_{iR'}=0}}.
        \end{equation}
\end{itemize}

\subsection{A nilpotent algebra of operators}

From subsection~\ref{subsec:lessons}, we see that $\hat{\mathbb N}$ and $\hat{\mathbb X}$ are important operators when constructing solutions of the Casimir recurrence relation \eqref{eq:covRecur}. It is therefore natural to study the commutators of $\hat{\mathbb N},\hat{\mathbb X},\hat{\mathbb F}$. It turns out that we only need three groups of operators to obtain a systematic solution of \eqref{eq:covRecur}:
\begin{subequations}\label{eq:covList}
\begin{align}
    \hat{\mathbb N}\mathcal Q&:=\sum_{a,i}\gamma_{ai}\langle\mathcal Q\rangle^{ai},&&\hat{\mathbb N}:\ \mathscr Q^{\Delta,J}\to\mathscr Q^{\Delta-1,J-1},\\
    \hat{\mathbb M}\mathcal Q&:=\sum_{a,i}\gamma_{aL}\gamma_{iR}\langle\mathcal Q\rangle^{ai},&&\hat{\mathbb M}:\ \mathscr Q^{\Delta,J}\to\mathscr Q^{\Delta-1,J-1},\\
    \nonumber\\
    \hat{\mathbb X}\mathcal Q&:=\sum_{\substack{a,b\\a\neq b}}\sum_{\substack{i,j\\i\neq j}}\gamma_{ab}\gamma_{ij}\{\langle\mathcal Q\rangle_{aj,bi}\}^{ab,ij},&&\hat{\mathbb X}:\ \mathscr Q^{\Delta,J}\to\mathscr Q^{\Delta+2,J},\\
    \hat{\mathbb Y}\mathcal Q&:=\sum_{\substack{a,b\\a\neq b}}\sum_{\substack{i,j\\i\neq j}}\gamma_{ab}\gamma_{ij}\{\langle\mathcal Q\rangle_{bi}\}^{ab,ij},&&\hat{\mathbb Y}:\ \mathscr Q^{\Delta,J}\to\mathscr Q^{\Delta+1,J-1},\\
    \hat{\mathbb Z}\mathcal Q&:=\sum_{\substack{a,b\\a\neq b}}\sum_{\substack{i,j\\i\neq j}}\gamma_{ab}\gamma_{ij}\{\mathcal Q\}^{ab,ij},&&\hat{\mathbb Z}:\ \mathscr Q^{\Delta,J}\to\mathscr Q^{\Delta,J-2},\\
    \nonumber\\
    \hat{\mathbb F}\mathcal Q&:=\sum_{\substack{a,b\\a\neq b}}\sum_{\substack{i,j\\i\neq j}}\gamma_{ai}\gamma_{bj}(\mathcal Q-\langle\mathcal Q\rangle_{aj,bi}^{ai,bj}),&&\hat{\mathbb F}:\ \mathscr Q^{\Delta,J}\to\mathscr Q^{\Delta,J},\\
    \hat{\mathbb G}\mathcal Q&:=\sum_{\substack{a,b\\a\neq b}}\sum_{\substack{i,j\\i\neq j}}\gamma_{ai}\gamma_{bj}(\langle\mathcal Q\rangle^{ai}-\langle\mathcal Q\rangle_{bi}^{ai,bj}),&&\hat{\mathbb G}:\ \mathscr Q^{\Delta,J}\to\mathscr Q^{\Delta-1,J-1}.
\end{align}
\end{subequations}
It is straightforward to compute their commutation relations. In particular, $\{\hat{\mathbb N},\hat{\mathbb X},\hat{\mathbb Y},\hat{\mathbb Z}\}$ close into a peculiar step-3 nilpotent algebra:
\begin{equation}
    [\hat{\mathbb N},\hat{\mathbb X}]=2\hat{\mathbb Y},\quad[\hat{\mathbb N},\hat{\mathbb Y}]=\hat{\mathbb Z},\quad[\hat{\mathbb N},\hat{\mathbb Z}]=0,\quad\hat{\mathbb X},\hat{\mathbb Y},\hat{\mathbb Z}\text{ commute}.
\end{equation}

The commutation relations involving $\hat{\mathbb M}$ are not so simple when acting on generic functions in $\mathscr Q^{\Delta,J}$. However, its action on factorizable functions is much simpler. In fact, $\{\hat{\mathbb M},\hat{\mathbb X},\hat{\mathbb Y},\hat{\mathbb Z}\}$ all preserve factorizability due to \eqref{eq:factorizability}. Specifically,
\begin{subequations}\label{eq:list2}
\begin{align}
    \hat{\mathbbm m}{\mathcal M}_{\mathcal L}&:=\sum_a\gamma_{aL}[{\mathcal M}_{\mathcal L}]^{aL}_{aL'},&&\hat{\mathbbm m}:\ \mathscr M_{\mathcal L}^{\Delta,J}\to\mathscr M_{\mathcal L}^{\Delta-1,J-1},\\
    \hat{\mathbbm x}{\mathcal M}_{\mathcal L}&:=\sum_{\substack{a,b\\a\neq b}}\gamma_{ab}[{\mathcal M}_{\mathcal L}]^{ab}_{aL,bL},&&\hat{\mathbbm x}:\ \mathscr M_{\mathcal L}^{\Delta,J}\to\mathscr M_{\mathcal L}^{\Delta+2,J},\\
    \hat{\mathbbm y}{\mathcal M}_{\mathcal L}&:=\sum_{\substack{a,b\\a\neq b}}\gamma_{ab}[{\mathcal M}_{\mathcal L}]^{ab}_{aL',bL},&&\hat{\mathbbm y}:\ \mathscr M_{\mathcal L}^{\Delta,J}\to\mathscr M_{\mathcal L}^{\Delta+1,J-1},\\
    \hat{\mathbbm z}{\mathcal M}_{\mathcal L}&:=\sum_{\substack{a,b\\a\neq b}}\gamma_{ab}[{\mathcal M}_{\mathcal L}]^{ab}_{aL',bL'},&&\hat{\mathbbm z}:\ \mathscr M_{\mathcal L}^{\Delta,J}\to\mathscr M_{\mathcal L}^{\Delta,J-2}.
\end{align}
\end{subequations}
Amazingly, these half-operators close into exactly the same step-3 nilpotent algebra:
\begin{equation}
    [\hat{\mathbbm m},\hat{\mathbbm x}]=2\hat{\mathbbm y},\quad[\hat{\mathbbm m},\hat{\mathbbm y}]=\hat{\mathbbm z},\quad[\hat{\mathbbm m},\hat{\mathbbm z}]=0,\quad\hat{\mathbbm x},\hat{\mathbbm y},\hat{\mathbbm z}\text{ commute}.
\end{equation}
This can also be understood from their action in position space. If ${\mathcal M}_{\mathcal L}\leftrightarrow G_{\mathcal L}$ are related through the Mellin transform \eqref{eq:spinMellin}, we have
\begin{subequations}
\begin{align}
    \hat{\mathbbm m}{\mathcal M}_{\mathcal L}\quad&\leftrightarrow\quad-(X_L\cdot\partial_{Z_L})G_{\mathcal L},\\
    \hat{\mathbbm x}{\mathcal M}_{\mathcal L}\quad&\leftrightarrow\quad-\frac12(\partial_{X_L}\cdot\partial_{X_L})G_{\mathcal L},\\
    \hat{\mathbbm y}{\mathcal M}_{\mathcal L}\quad&\leftrightarrow\quad-\frac12(\partial_{X_L}\cdot\partial_{Z_L})G_{\mathcal L},\\
    \hat{\mathbbm z}{\mathcal M}_{\mathcal L}\quad&\leftrightarrow\quad-\frac12(\partial_{Z_L}\cdot\partial_{Z_L})G_{\mathcal L}.
\end{align}
\end{subequations}
The nilpotent algebra is nothing but the algebra between these differential operators. In particular, $\hat{\mathbbm m}$ is precisely the operator needed in the gauge invariance condition \eqref{eq:singleGauge}, which reads $\hat{\mathbbm m}\mathcal M_{\mathcal L}=0$. This is crucial when we compute the action of $\hat{\mathbb M},\hat{\mathbb X},\hat{\mathbb Y},\hat{\mathbb Z}$, because $\hat{\mathbbm m}$ can always be commuted to the right to annihilate $\mathcal M_{\mathcal L}\mathcal M_{\mathcal R}$. For example,
\begin{align*}
    \hat{\mathbb M}\hat{\mathbb X}^m(\mathcal M_{\mathcal L}\mathcal M_{\mathcal R})&=(\hat{\mathbbm m}\hat{\mathbbm x}^m\mathcal M_{\mathcal L})(\hat{\mathbbm m}\hat{\mathbbm x}^m\mathcal M_{\mathcal R})=(2m\hat{\mathbbm y}\hat{\mathbbm x}^{m-1}\mathcal M_{\mathcal L})(2m\hat{\mathbbm y}\hat{\mathbbm x}^{m-1}\mathcal M_{\mathcal R})\\
    &=4m^2\hat{\mathbb Y}\hat{\mathbb X}^{m-1}(\mathcal M_{\mathcal L}\mathcal M_{\mathcal R}),\\
    \hat{\mathbb M}\hat{\mathbb X}\hat{\mathbb Y}(\mathcal M_{\mathcal L}\mathcal M_{\mathcal R})&=(\hat{\mathbbm m}\hat{\mathbbm x}\hat{\mathbbm y}\mathcal M_{\mathcal L})(\hat{\mathbbm m}\hat{\mathbbm x}\hat{\mathbbm y}\mathcal M_{\mathcal R})=((\hat{\mathbbm x}\hat{\mathbbm z}+2\hat{\mathbbm y}^2)\mathcal M_{\mathcal L})((\hat{\mathbbm x}\hat{\mathbbm z}+2\hat{\mathbbm y}^2)\mathcal M_{\mathcal R})\\
    &=\hat{\mathbb X}\hat{\mathbb Z}(\mathcal M_{\mathcal L}\mathcal M_{\mathcal R})+4\hat{\mathbb Y}^2(\mathcal M_{\mathcal L}\mathcal M_{\mathcal R})+2(\hat{\mathbbm x}\hat{\mathbbm z}\otimes\hat{\mathbbm y}^2+\hat{\mathbbm y}^2\otimes\hat{\mathbbm x}\hat{\mathbbm z})(\mathcal M_{\mathcal L}\mathcal M_{\mathcal R}).
\end{align*}
Note that cross terms such as $(\hat{\mathbbm x}\hat{\mathbbm z}\otimes\hat{\mathbbm y}^2+\hat{\mathbbm y}^2\otimes\hat{\mathbbm x}\hat{\mathbbm z})$ are inevitable, which suggests that the half-operators $\hat{\mathbbm x},\hat{\mathbbm y},\hat{\mathbbm z}$ are more fundamental objects.

For the operators $\hat{\mathbb F},\hat{\mathbb G}$, it turns out that all we need are their commutators with $\hat{\mathbb N}$:
\begin{equation}
    [\hat{\mathbb N},\hat{\mathbb F}]=2\hat{\mathbb G},\quad[\hat{\mathbb N},\hat{\mathbb G}]=0.
\end{equation}
The action of $\hat{\mathbb F}$ on factorizable functions is even simpler:
\begin{equation}
    \hat{\mathbb F}({\mathcal M}_{\mathcal L}{\mathcal M}_{\mathcal R})=0,
\end{equation}
and the action of $\hat{\mathbb G}$ on factorizable functions reads
\begin{equation}
    \hat{\mathbb G}({\mathcal M}_{\mathcal L}{\mathcal M}_{\mathcal R})=\left(\gamma_{LR}\hat{\mathbb N}-\hat{\mathbb M}\right)({\mathcal M}_{\mathcal L}{\mathcal M}_{\mathcal R}).
\end{equation}

\subsection{Systematic algorithm for the factorization formula}

With the operators defined above, we are ready to construct solutions $\mathcal Q_m$ to the Casimir recurrence \eqref{eq:covRecur}. The general strategy is to act on $\mathcal M_{\mathcal L}\mathcal M_{\mathcal R}$ with various operators. As we have seen, operators have simpler actions on factorizable functions, so we would like to preserve factorizability as much as possible. In practice, this means we should construct solutions by acting with $\hat{\mathbb M},\hat{\mathbb X},\hat{\mathbb Y},\hat{\mathbb Z}$ first and $\hat{\mathbb N}$ last. On general grounds, for a spin-$J$ exchange, we expect the residue $\mathcal Q_m$ to be a degree-$J$ polynomial with respect to the $\gamma_{ai}$ variables. Since each $\hat{\mathbb N}$ increases the degree of $\gamma_{ai}$ by one, this suggests the following ansatz:
\begin{equation}\label{eq:Qansatz}
    \mathcal Q_m=\left.\sum_{s=0}^J\hat{\mathbb N}^s\hat{\mathbbm P}_{s,m}(\mathcal M_{\mathcal L}\mathcal M_{\mathcal R})\right|_{\substack{\eta_{aL'}=0\\\eta_{iR'}=0}},
\end{equation}
where $\hat{\mathbbm P}_{s,m}(\hat{\mathbbm x},\hat{\mathbbm y},\hat{\mathbbm z}):\mathscr M_{\mathcal L}^{\Delta,J}\otimes\mathscr M_{\mathcal R}^{\Delta,J}\to\mathscr M_{\mathcal L}^{\Delta-J+s+2m,s}\otimes\mathscr M_{\mathcal R}^{\Delta-J+s+2m,s}$ is some polynomial in the half-operators. We do not need $\hat{\mathbbm m}$, because it can always be commuted to the end and annihilate $\mathcal M_{\mathcal L}\mathcal M_{\mathcal R}$.

Since $\hat{\mathbb F}$ vanishes and $\hat{\mathbb G}=\gamma_{LR}\hat{\mathbb N}-\hat{\mathbb M}$ on factorizable functions,
\begin{equation}
\begin{aligned}
    \hat{\mathbb F}\hat{\mathbb N}^s\hat{\mathbbm P}_{s,m}(\mathcal M_{\mathcal L}\mathcal M_{\mathcal R})&=2s\hat{\mathbb N}^{s-1}(\hat{\mathbb M}-\gamma_{LR}\hat{\mathbb N})\hat{\mathbbm P}_{s,m}(\mathcal M_{\mathcal L}\mathcal M_{\mathcal R})\\
    &=2s\hat{\mathbb N}^{s-1}\hat{\mathbb M}\hat{\mathbbm P}_{s,m}(\mathcal M_{\mathcal L}\mathcal M_{\mathcal R})-2s(\Delta-J+s+2m-1)\hat{\mathbb N}^s\hat{\mathbbm P}_{s,m}(\mathcal M_{\mathcal L}\mathcal M_{\mathcal R}).
\end{aligned}
\end{equation}
On the other hand,
\begin{equation}
    \hat{\mathbb X}\hat{\mathbb N}^s=\hat{\mathbb N}^s\hat{\mathbb X}-2s\hat{\mathbb N}^{s-1}\hat{\mathbb Y}+s(s-1)\hat{\mathbb N}^{s-2}\hat{\mathbb Z}.
\end{equation}
Since only $\hat{\mathbb N}$ breaks factorizability and depends on $\gamma_{ai}$ individually, \eqref{eq:covRecur} should hold for each degree of $\gamma_{ai}$. In other words, the coefficient of each $\hat{\mathbb N}^s$ should vanish individually. Putting everything together, we obtain
\begin{equation}\label{eq:doubleRecur}
    \begin{aligned}
        0=\,&\left(2(J-s)(\Delta-1+2m+s)-2m(2\Delta-d+2m)\right)\hat{\mathbbm P}_{s,m}(\mathcal M_{\mathcal L}\mathcal M_{\mathcal R})\\
        +\,&2(s+1)\hat{\mathbb M}\hat{\mathbbm P}_{s+1,m}(\mathcal M_{\mathcal L}\mathcal M_{\mathcal R})+\hat{\mathbb X}\hat{\mathbbm P}_{s,m-1}(\mathcal M_{\mathcal L}\mathcal M_{\mathcal R})\\
        -\,&2(s+1)\hat{\mathbb Y}\hat{\mathbbm P}_{s+1,m-1}(\mathcal M_{\mathcal L}\mathcal M_{\mathcal R})+(s+2)(s+1)\hat{\mathbb Z}\hat{\mathbbm P}_{s+2,m-1}(\mathcal M_{\mathcal L}\mathcal M_{\mathcal R}).
    \end{aligned}
\end{equation}
This is a double recurrence relation that determines $\hat{\mathbbm P}_{s,m}$ in terms of $\hat{\mathbbm P}_{\geq s,m}$ and $\hat{\mathbbm P}_{s,\leq m}$.

The double recurrence starts from $\hat{\mathbbm P}_{J,0}:\mathscr M_{\mathcal L}^{\Delta,J}\otimes\mathscr M_{\mathcal R}^{\Delta,J}\to\mathscr M_{\mathcal L}^{\Delta,J}\otimes\mathscr M_{\mathcal R}^{\Delta,J}$. Since $\hat{\mathbbm P}_{J,0}$ is a polynomial in $\hat{\mathbbm x},\hat{\mathbbm y},\hat{\mathbbm z}$, the only possibility is to take it to be a constant:
\begin{equation}
    \hat{\mathbbm P}_{J,0}=\mathcal K_{\Delta J}.
\end{equation}
Next, fixing $s=J$, we solve for $\hat{\mathbbm P}_{J,m}:\mathscr M_{\mathcal L}^{\Delta,J}\otimes\mathscr M_{\mathcal R}^{\Delta,J}\to\mathscr M_{\mathcal L}^{\Delta+2m,J}\otimes\mathscr M_{\mathcal R}^{\Delta+2m,J}$. Looking at how $\hat{\mathbbm x},\hat{\mathbbm y},\hat{\mathbbm z}$ change the support, we see that the only possibility is
\begin{equation}\label{eq:GJmansatz}
    \hat{\mathbbm P}_{J,m}=\mathcal N_{J,m}\hat{\mathbb X}^m.
\end{equation}
Plugging it into the recurrence relation, we obtain\footnote{These types of recurrence relations can be easily solved using \texttt{RSolve[]} in Mathematica. In practice, a better method is to plug in many specific values of $m$ and use \texttt{FindSequenceFunction[]} to find $\mathcal N_{J,m}$.}
\begin{equation}\label{eq:GJmresult}
    \left(-2m(2\Delta-d+2m)\right)\mathcal N_{J,m}+\mathcal N_{J,m-1}=0\ \implies\ \mathcal N_{J,m}=\frac{\mathcal K_{\Delta J}}{4^mm!(1-\frac d2+\Delta)_m}.
\end{equation}
Although we have only solved the $s=J$ part in the ansatz \eqref{eq:Qansatz}, this is actually the complete factorization formula for scalar ($J=0$) exchanges! And indeed, choosing the normalization $\mathcal K_{\Delta0}=\kappa_{\Delta0}$ precisely recovers \eqref{eq:scalarFac}.

We now continue to $s=J-1$. The only polynomial for $\hat{\mathbbm P}_{J-1,m}:\mathscr M_{\mathcal L}^{\Delta,J}\otimes\mathscr M_{\mathcal R}^{\Delta,J}\to\mathscr M_{\mathcal L}^{\Delta-1+2m,J-1}\otimes\mathscr M_{\mathcal R}^{\Delta-1+2m,J-1}$ is
\begin{equation}
    \hat{\mathbbm P}_{J-1,m}=\mathcal N_{J-1,m}\hat{\mathbb X}^{m-1}\hat{\mathbb Y}.
\end{equation}
Plugging into the recurrence relation, we have
\begin{equation}
\begin{aligned}
    &\left(2(\Delta+J-2+2m)-2m(2\Delta-d+2m)\right)\mathcal N_{J-1,m}+\mathcal N_{J-1,m-1}\\
    =\,&{-}2J(4m^2\mathcal N_{J,m}-\mathcal N_{J,m-1}).
\end{aligned}
\end{equation}
Imposing the boundary condition $\mathcal N_{J-1,-1}=0$, we obtain
\begin{equation}
    \mathcal N_{J-1,m}=\frac{\mathcal K_{\Delta J}}{4^mm!(1-\frac d2+\Delta)_m}\frac{2Jm(d-2\Delta)}{\Delta-J-d+2}.
\end{equation}
An important observation is that although we only demanded $\mathcal N_{J-1,-1}=0$, it turns out that $\mathcal N_{J-1,0}=0$ too. This is a non-trivial consistency check, because $\mathcal N_{J-1,0}$ would be the coefficient of $\hat{\mathbb X}^{-1}\hat{\mathbb Y}$ which contains negative powers. Also, we remark that with the $s=J$ and $s=J-1$ results at hand, we are able to recover the complete factorization formula for spin $J=1$ exchanges. Specifically, choosing $\mathcal K_{\Delta1}=2\kappa_{\Delta1}$ recovers (97) of \cite{Goncalves:2014rfa}:
\begin{equation}
    J=1:\quad\mathcal Q_m=\frac{\mathcal K_{\Delta1}}{4^mm!(1-\frac d2+\Delta)_m}\left.\left(\hat{\mathbb N}\hat{\mathbb X}^m+\frac{2m(d-2\Delta)}{\Delta-d+1}\hat{\mathbb X}^{m-1}\hat{\mathbb Y}\right)(\mathcal M_{\mathcal L}\mathcal M_{\mathcal R})\right|_{\substack{\eta_{aL'}=0\\\eta_{iR'}=0}}.
\end{equation}

For $s=J-2$ and $\hat{\mathbbm P}_{J-2,m}:\mathscr M_{\mathcal L}^{\Delta,J}\otimes\mathscr M_{\mathcal R}^{\Delta,J}\to\mathscr M_{\mathcal L}^{\Delta-2+2m,J-2}\otimes\mathscr M_{\mathcal R}^{\Delta-2+2m,J-2}$, we now have three possibilities:
\begin{equation}
    \hat{\mathbbm P}_{J-2,m}=\mathcal N_{J-2,m}^{(1)}\hat{\mathbb X}^{m-2}\hat{\mathbb Y}^2+\mathcal N_{J-2,m}^{(2)}\hat{\mathbb X}^{m-1}\hat{\mathbb Z}+\mathcal N_{J-2,m}^{(3)}\hat{\mathbb X}^{m-2}\frac{\hat{\mathbbm x}\hat{\mathbbm z}\otimes\hat{\mathbbm y}^2+\hat{\mathbbm y}^2\otimes\hat{\mathbbm x}\hat{\mathbbm z}}2.
\end{equation}
Note that we require $\{\hat{\mathbbm x},\hat{\mathbbm y},\hat{\mathbbm z}\}$ to appear symmetrically, because the factorization formula should not depend on which side we call left. Plugging into the recurrence relation, we obtain a recurrence relation for each of the three structures. Demanding $\mathcal N_{J-2,-1}^{(1,2,3)}=0$, we obtain a unique solution:
\begin{subequations}
\begin{align}
    \frac{\mathcal N_{J-2,m}^{(1)}}{\mathcal N_{J,m}}&=\frac{2J(J-1)m(m-1)(2\Delta-d)(2\Delta-d+2)}{(\Delta-J-d+2)(\Delta-J-d+3)},\\
    \frac{\mathcal N_{J-2,m}^{(2)}}{\mathcal N_{J,m}}&=J(J-1)m\Big[1+\frac{2m}{\Delta-J-d+2}-\frac{2 (\Delta +J-3) (\Delta -J-d+m+2)}{(2 J+d-4) (\Delta - J -d +3)}\Big],\\
    \frac{\mathcal N_{J-2,m}^{(3)}}{\mathcal N_{J,m}}&=-\frac{4J(J-1)m(m-1)(2\Delta-d)}{(\Delta-J-d+2)(\Delta-J-d+3)}.
\end{align}
\end{subequations}
Again, we see that although we only demanded $\mathcal N_{J-2,-1}^{(1,2,3)}=0$, we still obtain $\mathcal N_{J-2,m}^{(1,3)}=0$ for $m=0,1$ and $\mathcal N_{J-2,m}^{(2)}=0$ for $m=0$, consistent with the fact that $\hat{\mathbb X}$ has to appear with non-negative powers. Also, with the $s=J,J-1,J-2$ results at hand, we are able to recover the complete factorization formula for spin $J=2$ exchanges. Specifically, choosing $\mathcal K_{\Delta2}=\kappa_{\Delta2}$ recovers (107) of \cite{Goncalves:2014rfa}, with the precise dictionary given by 
\begin{subequations}
    \begin{gather}
    L_m^{ab}=\frac1{2^mm!}[\hat{\mathbbm x}^m\mathcal M_{\mathcal L}]^{aL,bL}_{aL',bL'}\Big|_{\eta=0},\quad\dot L_m^a=\frac1{2^{m-1}(m-1)!}[\hat{\mathbbm x}^{m-1}\hat{\mathbbm y}\mathcal M_{\mathcal L}]^{aL}_{aL'}\Big|_{\eta=0},\\[1mm]
    \ddot L_m=\frac1{2^{m-1}(m-1)!}\hat{\mathbbm m}\hat{\mathbbm x}^{m-1}\hat{\mathbbm y}\mathcal M_{\mathcal L}\Big|_{\eta=0},\quad\widetilde L_m=\frac1{2^{m-1}(m-1)!}\hat{\mathbbm x}^{m-1}\hat{\mathbbm z}\mathcal M_{\mathcal L}\Big|_{\eta=0},\\[1mm]
    h_m^{(3)}=4^{m-2}((m-2)!)^2\mathcal N_{2-2,m}^{(1)},\quad h_m^{(3)}+2h_m^{(4)}+h_m^{(5)}=4^{m-1}((m-1)!)^2\mathcal N_{2-2,m}^{(2)},\\[1mm]
    h_m^{(3)}+h_m^{(4)}=4^{m-2}(m-1)!(m-2)!\mathcal N_{2-2,m}^{(3)}.
\end{gather}
\end{subequations}

Using the method described above, we can systematically solve for all $\hat{\mathbbm P}_{s,m}$. If we require the half-operators to map $\mathscr M_{\mathcal L/\mathcal R}^{\Delta,J}\to\mathscr M_{\mathcal L/\mathcal R}^{\Delta-J+s+2m,s}$, it is not hard to show that there are exactly
\begin{equation*}
    \left\lceil\frac{J-s+1}2\right\rceil=1,1,2,2,3,\cdots
\end{equation*}
choices of half-operators. More explicitly, starting at $\hat{\mathbbm x}^m$ for $s=J$, whenever we decrease $s$ by one, replace one $\hat{\mathbbm x}$ by $\hat{\mathbbm y}$ and then replace $\hat{\mathbbm y}^2\mapsto\hat{\mathbbm x}\hat{\mathbbm z}$ any number of times:
\begin{equation*}
    \begin{array}{c|l|c}
        J-s & \mathscr M_{\mathcal L/\mathcal R}^{\Delta,J}\to\mathscr M_{\mathcal L/\mathcal R}^{\Delta-J+s+2m,s} & \text{number of half-structures} \\
        \hline\hline
        0 & \hat{\mathbbm x}^m & 1 \\
        \hline
        1 & \hat{\mathbbm x}^{m-1}\hat{\mathbbm y} & 1\\
        \hline
        2 & \hat{\mathbbm x}^{m-2}\hat{\mathbbm y}^2,\ \hat{\mathbbm x}^{m-1}\hat{\mathbbm z}, & 2 \\
        \hline
        3 & \hat{\mathbbm x}^{m-3}\hat{\mathbbm y}^3,\ \hat{\mathbbm x}^{m-2}\hat{\mathbbm y}\hat{\mathbbm z}, & 2 \\
        \hline
        4 & \hat{\mathbbm x}^{m-4}\hat{\mathbbm y}^4,\ \hat{\mathbbm x}^{m-3}\hat{\mathbbm y}^2\hat{\mathbbm z},\ \hat{\mathbbm x}^{m-2}\hat{\mathbbm z}^2 & 3 \\
        \hline
        \vdots &  & \vdots
    \end{array}
\end{equation*}
The possible structures in $\hat{\mathbb P}_{s,m}$ can be found by pairing the left and right half-operators symmetrically, leading to a total number
\begin{equation*}
    \frac12\times\left\lceil\frac{J-s+1}2\right\rceil\times\left(\left\lceil\frac{J-s+1}2\right\rceil+1\right)=1,1,3,3,6,6,10,\cdots
\end{equation*}
For any spin-$J$ exchange, only those structures with $J-s\leq J$ are needed. As a result, there are $1$, $2=1+1$, $5=2+3$, $8=5+3$, $14=8+6$ terms in total in the spin-0,1,2,3,4 factorization formulas. We provide the explicit factorization formulas up to spin-4 in an ancillary file.

As a final check, consider $\hat{\mathbbm P}_{s,m=0}:\mathscr M_{\mathcal L}^{\Delta,J}\otimes\mathscr M_{\mathcal R}^{\Delta,J}\to\mathscr M_{\mathcal L}^{\Delta-J+s,s}\otimes\mathscr M_{\mathcal R}^{\Delta-J+s,s}$. The only polynomial of $\hat{\mathbbm x},\hat{\mathbbm y},\hat{\mathbbm z}$ that yields something of this form is a constant, which then implies that $s=J$. In other words, $\mathcal N_{J,0}=\mathcal K_{\Delta J}$ is sufficient for the primary factorization formula. By choosing
\begin{equation}
    \mathcal K_{\Delta J}=\frac{2^J\kappa_{\Delta J}}{(J!)^2},
\end{equation}
we obtain
\begin{equation}
    \mathcal Q_0=\frac{2^J\kappa_{\Delta J}}{(J!)^2}\hat{\mathbb N}^J(\mathcal M_{\mathcal L}\otimes\mathcal M_{\mathcal R}),
\end{equation}
which proves the conjecture \eqref{eq:primaryFac}.

\subsection{A collection of factorization formulas}\label{subsec:facform}

 We collect all factorization formulas derived in the operator language here for the reader's convenience.
\begin{itemize}
    \item { Scalar factorization:}
    \begin{equation}
        \mathcal Q_m=\frac{\mathcal K_{\Delta0}}{4^mm!(1-\frac d2+\Delta)_m}\left.\hat{\mathbb X}^m \circ(\mathcal M_{\mathcal L}\mathcal M_{\mathcal R})\right|_{\substack{\eta_{aL'}=0\\\eta_{iR'}=0}}.
    \end{equation}
    \item { Spin-1 factorization:}
    \begin{equation}
        \mathcal Q_m=\frac{\mathcal K_{\Delta1}}{4^mm!(1-\frac d2+\Delta)_m}\left.\left(\hat{\mathbb N}\hat{\mathbb X}^m+\frac{2m(d-2\Delta)}{\Delta-d+1}\hat{\mathbb X}^{m-1}\hat{\mathbb Y}\right)(\mathcal M_{\mathcal L}\mathcal M_{\mathcal R})\right|_{\substack{\eta_{aL'}=0\\\eta_{iR'}=0}}.
    \end{equation}
    \item { Spin-2 factorization:}
    \begin{equation}
    \begin{aligned}
        \mathcal Q_m = \frac{\mathcal K_{\Delta 2}}{4^mm!(1-\frac d2+\Delta)_m} &\left[ \hat{\mathbb N}^2\hat{\mathbb X}^m + A_1\, \hat{\mathbb N}\hat{\mathbb X}^{m-1}\hat{\mathbb Y} + A_2\, \hat{\mathbb X}^{m-2}\hat{\mathbb Y}^2 +A_3\, \hat{\mathbb X}^{m-1}\hat{\mathbb Z} \right. \\
        &\left.\left. +A_4\, \hat{\mathbb X}^{m-2}(\hat{\mathbbm x}\hat{\mathbbm z}\!\otimes\!\hat{\mathbbm y}^2\!+\!\hat{\mathbbm y}^2\!\otimes\!\hat{\mathbbm x}\hat{\mathbbm z}) \right](\mathcal M_{\mathcal L}\mathcal M_{\mathcal R})\right|_{\substack{\eta_{aL'}=0\\\eta_{iR'}=0}},
    \end{aligned}
    \end{equation}
    with
    \begin{subequations}
    \begin{align}
        A_1 &= -\frac{4m(2\Delta-d)}{\Delta-d},\\
        A_2 &=\frac{4m(m-1)(2\Delta-d)(2\Delta-d+2)}{(\Delta-d)(\Delta-d+1)},\\
        A_3 &=2m\left[1+\frac{2m}{\Delta-d}-\frac{2 (\Delta-1) (\Delta-d+m)}{d(\Delta -d +1)}\right],\\
        A_4 &=-\frac{4m(m-1)(2\Delta-d)}{(\Delta-d)(\Delta-d+1)}.
    \end{align}
    \end{subequations}
\end{itemize}
Here, the normalization $\mathcal K_{\Delta J}$ is defined by 
\begin{equation}
    \mathcal K_{\Delta J}=\frac{(-1)^{J-1}}{(J!)^2}2(\Delta+J-1)\Gamma(\Delta-1).
\end{equation}
The definitions of these shift operators can be found in \eqref{eq:shifts}, \eqref{eq:covList} and \eqref{eq:list2}.

\subsection{Generalization to mixed-symmetry tensors}

Finally, we briefly comment on how to generalize to mixed-symmetry tensors. The strategy is exactly the same, although the resulting factorization formulas will be significantly more complicated. For example, if the exchanged operator $L/R$ has spin-$(J_{(1)},J_{(2)})$, we simply introduce more shift half-operators like $\{\hat{\mathbbm m},\hat{\mathbbm x},\hat{\mathbbm y},\hat{\mathbbm z}\}$:
\begin{subequations}
\begin{gather}
    \hat{\mathbbm m}_{10\mathcal L}\leftrightarrow-(X_L\cdot\partial_{Z_L^{(1)}}),\quad\hat{\mathbbm m}_{21\mathcal L}\leftrightarrow-(Z_L^{(1)}\cdot\partial_{Z_L^{(2)}}),\quad\hat{\mathbbm m}_{20\mathcal L}\leftrightarrow-(X_L\cdot\partial_{Z_L^{(2)}}),\\
    \hat{\mathbbm y}_{00\mathcal L}\leftrightarrow-\frac12(\partial_{X_L}\cdot\partial_{X_L}),\ \hat{\mathbbm y}_{01\mathcal L}\leftrightarrow-\frac12(\partial_{X_L}\cdot\partial_{Z_L^{(1)}}),\ \cdots,\ \hat{\mathbbm y}_{22\mathcal L}\leftrightarrow-\frac12(\partial_{Z_L^{(2)}}\cdot\partial_{Z_L^{(2)}}).
\end{gather}
\end{subequations}
These satisfy a slightly more complicated nilpotent algebra, as can be derived from position space, and $\hat{\mathbbm m}_{10\mathcal L},\hat{\mathbbm m}_{21\mathcal L},\hat{\mathbbm m}_{20\mathcal L}$ all annihilate $\mathcal M_{\mathcal L}$ due to gauge invariance. These shift half-operators can be viewed as the action of shift operators on factorizable functions. For example,
\begin{subequations}
\begin{align}
    \hat{\mathbb M}_{20}\mathcal Q&:=\sum_{a,i}\gamma_{aL}\gamma_{iR}[\mathcal Q]^{aL,iR,ai,LR}_{aL'',iR''},\\
    \hat{\mathbb M}_{21}\mathcal Q&:=\sum_{a,i}\gamma_{aL}\gamma_{iR}[\mathcal Q]^{aL',iR'}_{aL'',iR''},\\
    \vdots\nonumber\\
    \hat{\mathbb Y}_{02}\mathcal Q&:=\sum_{\substack{a,b\\a\neq b}}\sum_{\substack{i,j\\i\neq j}}\gamma_{ab}\gamma_{ij}[\mathcal Q]^{ab,ij}_{bL,iR,bi,LR,aL'',jR''},\\
    \hat{\mathbb Y}_{12}\mathcal Q&:=\sum_{\substack{a,b\\a\neq b}}\sum_{\substack{i,j\\i\neq j}}\gamma_{ab}\gamma_{ij}[\mathcal Q]^{ab,ij}_{bL',iR',aL'',jR''},\\
    \vdots\nonumber
\end{align}
\end{subequations}
Similarly, we can define $\hat{\mathbb N}_{10},\hat{\mathbb N}_{21},\hat{\mathbb N}_{20}$ that do not preserve factorizability:
\begin{subequations}
\begin{align}
    \hat{\mathbb N}_{20}\mathcal Q&:=\sum_{a,i}\gamma_{ai}[\mathcal Q]^{aL,iR,ai,LR}_{aL'',iR''},\\
    \hat{\mathbb N}_{21}\mathcal Q&:=\sum_{a,i}\gamma_{ai}[\mathcal Q]^{aL',iR'}_{aL'',iR''},\\
    \vdots\nonumber
\end{align}
\end{subequations}
It can be checked that $\hat{\mathbb N}$ and $\hat{\mathbb Y}$ again form the same nilpotent algebra as $\hat{\mathbbm m}_{\mathcal L}$ and $\hat{\mathbbm y}_{\mathcal L}$. In order to work out the factorization formula, simply use the ansatz
\begin{equation}
    \mathcal Q_m=\sum_{\substack{s_{10},s_{21},s_{20}\geq0\\s_{21}+s_{20}=J_{(2)}\\s_{10}-s_{21}=J_{(1)}}}\hat{\mathbb N}_{10}^{s_{10}}\hat{\mathbb N}_{21}^{s_{21}}\hat{\mathbb N}_{20}^{s_{20}}\hat{\mathbbm P}_{s_{10},s_{21},s_{20},m}(\mathcal M_{\mathcal L}\mathcal M_{\mathcal R})\Big|_{\substack{\eta_{aL'}=\eta_{aL''}=0\\\eta_{iR'}=\eta_{iR''}=0}}
\end{equation}
to solve the Casimir recurrence equation.

\section{Applications}\label{sec:boots}

\subsection{Bootstrapping $\langle\mathcal V_\Delta\mathcal V_\Delta\mathcal V_\Delta\rangle$: the AdS on-shell method}\label{subsec:gluon3gen}

We are now ready to put our definition of the spinning Mellin amplitude to use. As a first example, we will bootstrap the three-point function of vector operators $\mathcal V_\Delta(X,Z)$ with scaling dimension $\Delta$:
\begin{equation*}
    \langle\mathcal V_\Delta(X_1,Z_1)\mathcal V_\Delta(X_2,Z_2)\mathcal V_\Delta(X_3,Z_3)\rangle.
\end{equation*}
In the AdS/CFT correspondence, $\mathcal{V}_\Delta$ is generally dual to a massive spin-1 particle in the bulk. When $\Delta = d-1$, this operator becomes a conserved current, and the corresponding bulk particle is a gluon.

By definition, the Mellin amplitude depends on 3 continuous variables $\gamma_{12},\gamma_{13},\gamma_{23}$ and 9 discrete variables $\eta_{1'2},\eta_{1'3},\eta_{2'1},\eta_{2'3},\eta_{3'1},\eta_{3'2},\eta_{1'2'},\eta_{1'3'},\eta_{2'3'}$, subject to the constraints:
\begin{subequations}
\begin{align}
    \gamma_{21}-\eta_{2'1}+\gamma_{31}-\eta_{3'1}&=\Delta,&\eta_{21'}+\eta_{2'1'}+\eta_{31'}+\eta_{3'1'}&=1,\\
    \gamma_{12}-\eta_{1'2}+\gamma_{32}-\eta_{3'2}&=\Delta,&\eta_{12'}+\eta_{1'2'}+\eta_{32'}+\eta_{3'2'}&=1,\\
    \gamma_{13}-\eta_{1'3}+\gamma_{23}-\eta_{2'3}&=\Delta,&\eta_{13'}+\eta_{1'3'}+\eta_{23'}+\eta_{2'3'}&=1.
\end{align}
\end{subequations}
We can solve for $\gamma_{12},\gamma_{13},\gamma_{23}$ in terms of the discrete $\eta$-variables, and after requiring the Mellin amplitude to have the form \eqref{eq:translate2}, with multi-linear dependence on the labels $1',2',3'$, we obtain the following ansatz:
\begin{equation}
    \begin{aligned}
        \mathcal M&=c_1\eta_{1'2'}\eta_{3'1}+c_2\eta_{1'2'}\eta_{3'2}+c_3\eta_{2'3'}\eta_{1'2}+c_4\eta_{2'3'}\eta_{1'3}+c_5\eta_{3'1'}\eta_{2'3}+c_6\eta_{3'1'}\eta_{2'1}\\
        &+c_7\eta_{1'2}\eta_{2'3}\eta_{3'1}+c_8\eta_{1'2}\eta_{2'3}\eta_{3'2}+c_9\eta_{1'2}\eta_{2'1}\eta_{3'1}+c_{10}\eta_{1'2}\eta_{2'1}\eta_{3'2}\\
        &+c_{11}\eta_{1'3}\eta_{2'3}\eta_{3'1}+c_{12}\eta_{1'3}\eta_{2'3}\eta_{3'2}+c_{13}\eta_{1'3}\eta_{2'1}\eta_{3'1}+c_{14}\eta_{1'3}\eta_{2'1}\eta_{3'2}.
    \end{aligned}
\end{equation}
The Mellin amplitude satisfies the gauge-invariance condition \eqref{eq:gaugeFinal} of all three external particles. For example, consider the gauge-invariance of particle 3:
\begin{equation}\label{eq:gi3}
    \gamma_{13}[\mathcal M]^{13}_{13'}-\eta_{1'3}[\mathcal M]^{1'3}_{1'3'}+\gamma_{23}[\mathcal M]^{23}_{23'}-\eta_{2'3}[\mathcal M]^{2'3}_{2'3'}=0,
\end{equation}
which holds on a different support with respect to the operator 3 due to the shifts:
\begin{align}
    \gamma_{13}-\eta_{1'3}+\gamma_{23}-\eta_{2'3}&=\Delta-1,&\eta_{13'}+\eta_{1'3'}+\eta_{23'}+\eta_{2'3'}&=0.
\end{align}
Since the discrete $\eta$ variables are non-negative, this corresponds to 5 lattice points. Solving \eqref{eq:gi3} on these lattice points, we obtain
\begin{equation}
    \begin{gathered}
        c_2=-c_1,\quad c_{10}=-c_9,\quad c_5=-c_4+\frac{\Delta+1}2c_{11}+\frac{\Delta+1}2c_{12},\\
        c_7=\frac2{\Delta-1}c_3-\frac{\Delta+1}{\Delta-1}c_8,\quad c_{14}=\frac2{\Delta-1}c_6-\frac{\Delta+1}{\Delta-1}c_{13}.
    \end{gathered}
\end{equation}
Similarly, we can impose the gauge-invariance conditions of particles 1 and 2. Together, they are able to fix 10 of the 14 unknown coefficients in the ansatz, which leads to
\begin{equation}\label{eq:g3aftergauge}
    \begin{aligned}
        \mathcal M&=\frac{c_8}2(\eta_{1'2}-\eta_{1'3})((\Delta+1)\eta_{2'3'}+2\eta_{2'3}\eta_{3'2})\\
        &+\frac{c_{11}}2(\eta_{2'3}-\eta_{2'1})((\Delta+1)\eta_{1'3'}+2\eta_{1'3}\eta_{3'1})\\
        &+\frac{c_9}2(\eta_{3'1}-\eta_{3'2})((\Delta+1)\eta_{1'2'}+2\eta_{1'2}\eta_{2'1})\\
        &+\frac{c_7}2((\Delta-1)(\eta_{1'2'}\eta_{3'1}-\eta_{1'2'}\eta_{3'2}+\eta_{2'3'}\eta_{1'2}-\eta_{2'3'}\eta_{1'3}+\eta_{3'1'}\eta_{2'3}-\eta_{3'1'}\eta_{2'1})\\
        &\quad\quad+2(\eta_{1'2}\eta_{2'3}\eta_{3'1}-\eta_{1'3}\eta_{2'1}\eta_{3'2})).
    \end{aligned}
\end{equation}

Vector operators saturating the unitarity bound $\Delta=d-1$ are conserved currents. In the embedding formalism, the conservation condition translates to \cite{Costa:2011mg,Goncalves:2014rfa}:
\begin{equation}
    \forall1\leq k\leq3:\quad\frac\partial{\partial X_k}\cdot\frac\partial{\partial Z_k}\langle\mathcal V_\Delta(X_1,Z_1)\mathcal V_\Delta(X_2,Z_2)\mathcal V_\Delta(X_3,Z_3)\rangle=0.
\end{equation}
In Mellin space this becomes an equation on the Mellin amplitude $\mathcal M$. For example, conservation of operator 3 reads\footnote{This is in fact $\hat{\mathbbm y}\mathcal M=0$ if we treat 3 as the ``exchanged operator'' $L$.
For higher-spin conserved currents with spin $J>1$, define $\hat{\mathbbm w}\mathcal M:=\sum_c\gamma_{cL'}[\mathcal M]^{cL'}_{cL}\leftrightarrow-(Z_L\cdot\partial_{X_L})G$.
The conservation condition gives $[(d-4+2J)\hat{\mathbbm y}+\hat{\mathbbm w}\hat{\mathbbm z}]\mathcal M=0$ (eq.~(59) of~\cite{Goncalves:2014rfa}).}
\begin{equation}\label{eq:ub3}
    \sum_{\substack{a,b\in\{1,1',2,2'\}\\a\neq b}}\gamma_{ab}[\mathcal M]^{ab}_{a3',b3}=0,
\end{equation}
which holds on the following support with respect to the operator 3 due to the shifts:
\begin{align}
    \gamma_{13}-\eta_{1'3}+\gamma_{23}-\eta_{2'3}&=\Delta+1,&\eta_{13'}+\eta_{1'3'}+\eta_{23'}+\eta_{2'3'}&=0.
\end{align}
Since the discrete $\eta$ variables are non-negative, this corresponds to 5 lattice points. Solving \eqref{eq:ub3} on these lattice points, we obtain $c_8=c_{11}$ in \eqref{eq:g3aftergauge}. Similarly, we can impose conservation of all three operators. Together, they fix $c_8=c_9=c_{11}$. In the end, we obtain
\begin{equation}\label{eq:g3generic}
    \begin{aligned}
        \mathcal M&=c\Big(\eta_{1'2'}(\eta_{3'1}-\eta_{3'2})+\eta_{2'3'}(\eta_{1'2}-\eta_{1'3})+\eta_{3'1'}(\eta_{2'3}-\eta_{2'1})\\
        &\qquad+\frac2{\Delta-1}(\eta_{1'2}\eta_{2'3}\eta_{3'1}-\eta_{1'3}\eta_{2'1}\eta_{3'2})\Big)\\
        &-c'\Big((\eta_{1'2}-\eta_{1'3})(\eta_{2'3}-\eta_{2'1})(\eta_{3'1}-\eta_{3'2})\\
        &\qquad+\frac2{\Delta-1}(\eta_{1'2}\eta_{2'3}\eta_{3'1}-\eta_{1'3}\eta_{2'1}\eta_{3'2})\Big).
    \end{aligned}
\end{equation}
Several remarks are in order:
\begin{itemize}
    \item The spacetime dimension $d$ does not appear in \eqref{eq:g3aftergauge}, which manifests the view that the Mellin formalism is a $d$-independent representation of conformal field theories~\cite{Mack:2009mi}. 
    \item Although not imposed manually, the kinematic dependence of \eqref{eq:g3generic} is automatically antisymmetric under permuting the external operators $1,2,3$. Since the correlator itself is permutation invariant, the coupling constants $c,c'$ must be antisymmetric as well. This is expected, as the coupling constants are nothing but the structure constants $f^{abc}$ of a global symmetry group, under which these Noether currents transform in the adjoint representation.
    \item The fact that there are two independent solutions \eqref{eq:g3generic} for the three-point gluon amplitude is not surprising. This is because the effective bulk Lagrangian can have $F^2$ and $F^3$ interactions\footnote{Moreover, the four solutions in \eqref{eq:g3aftergauge} could be thought of as arising from $F^3$ and $A^\mu(A^\nu\overleftrightarrow\nabla_\mu A_\nu)$.}. The precise values of $c,c'$ depend on the theory considered. For example, in pure Yang-Mills theory, (6.10) of~\cite{Paulos:2011ie}\footnote{In eq.(5.4) and eq.(6.13) of~\cite{Paulos:2011ie}, the correct interpretation of $P_{ij}$ should be $P_i\cdot P_j$ instead of $-2P_i\cdot P_j$ as defined in eq.(2.3). This could be checked either by actually performing the $D^{M_1A}D^{M_2B}D^{M_3C}$ differentiation in eq.(6.10), or against the explicit result eq.(2.13) of~\cite{Baumann:2024ttn} in $d=3$.} implies $c'/c=\frac{1}{2d-3}$.
    \item Note that up to the correction terms proportional to $\frac2{\Delta-1}$, the AdS amplitude \eqref{eq:g3generic} takes a form similar to the flat-space amplitude
    \begin{equation}
        \begin{aligned}
            \mathcal M^{\rm flat}&=g_{F^2}\left((\mathtt e_1\cdot\mathtt e_2)\mathtt e_3\cdot(\mathtt k_1-\mathtt k_2)+\text{cyclic}\right)\\
            &+g_{F^3}\left(\mathtt e_1\cdot(\mathtt k_2-\mathtt k_3)\mathtt e_2\cdot(\mathtt k_3-\mathtt k_1)\mathtt e_3\cdot(\mathtt k_1-\mathtt k_2)\right).
        \end{aligned}
    \end{equation}
    This suggests the identification of $\eta_{p'q}\sim\mathtt e_p\cdot\mathtt k_q$ and $\eta_{p'q'}\sim\mathtt e_p\cdot\mathtt e_q$ in the flat-space limit.
\end{itemize}

\subsection{Case study: $\mathcal N=2$ theory of supergluons}\label{subsec:supergluons}

Our definition of (multi-)spinning Mellin amplitudes greatly facilitates the study of spinning correlators in specific theories. In this subsection, we showcase this by considering supergluons in $\mathrm{AdS}_5\times \mathrm{S}^3$~\cite{Fayyazuddin:1998fb,Aharony:1998xz,Karch:2002sh} (see \cite{Alday:2021odx,Zhou:2021gnu,Alday:2021ajh,Alday:2022lkk,Bissi:2022wuh,Huang:2023oxf,Alday:2023kfm,Huang:2023ppy,Cao:2023cwa,Huang:2024dck,Alday:2024yax,Alday:2024ksp,Cao:2024bky,Huang:2024rxr,Huang:2024dxr,Wang:2025pjo,Wang:2025owf} for recent progress in bootstrapping this theory). From the boundary point of view, this theory is an $\mathcal N=2$ SCFT in $d=4$ dimensions. We restrict our attention to the half-BPS supermultiplet, where the superprimary operator
\begin{equation}
    \mathcal O^a(x,v)=\mathcal O^{a;\alpha_1\alpha_2}(x)v^{\beta_1}v^{\beta_2}\epsilon_{\alpha_1\beta_1}\epsilon_{\alpha_2\beta_2}
\end{equation}
is a scalar with dimension $\Delta=2$. Here, $a=1,\cdots,\dim G_F$ is the adjoint index of a flavor group $G_F$, which we take to be $G_F=SU(N_f)$, and $v^\beta$ is an auxiliary $SU(2)_R$ spinor which ensures $\mathcal O^{a;\alpha_1\alpha_2}$ lives in the R-spin-1 representation. Another important primary operator in the supermultiplet is the conserved current $\mathcal J_\mu^a(x)$ of the flavor group $G_F$, which is an $SU(2)_R$ singlet and has dimension $\Delta=3$. This theory is dual to an $\mathcal N=1$ super-Yang-Mills theory on ${\rm AdS}_5\times S^3$ in the bulk, where $\mathcal O^a$ corresponds to a scalar field (supergluon) and $\mathcal J_\mu^a$ corresponds to a gauge field (gluon). When considering the correlators of $\mathcal O$ and $\mathcal J$ at tree level\footnote{More precisely, we are considering the probe limit, which is the leading order in $N_c\to\infty$ while keeping $N_f$ finite. In this case, gravity decouples and we are essentially scattering (super)gluons on the fixed AdS background.}, the only single-particle states propagating in the bulk are $\mathcal O$ and $\mathcal J$, and there are no contributions from other operators in the factorization. 

Denoting the two operators collectively by $\mathcal X=\mathcal O,\mathcal J$, we shall study their connected correlators. At tree level in the bulk, we can perform color decomposition in the same way as in flat space~\cite{DelDuca:1999rs}:
\begin{equation}
    \langle\mathcal X^{a_1}(X_1,Z_1)\cdots\mathcal X^{a_N}(X_N,Z_N)\rangle=\sum_{\sigma\in S_{N-1}}\mathop{\rm tr}(T^{a_1}T^{a_{2^\sigma}}\cdots T^{a_{N^\sigma}})\langle\mathcal X_1\mathcal X_{2^\sigma}\cdots\mathcal X_{N^\sigma}\rangle,
\end{equation}
where $\sigma$ denotes a permutation of the labels $\{2,\cdots,N\}$. In particular, if all $N$ operators are of the same type, we will denote $G^\mathcal X_{1\sigma}:=\langle\mathcal X_1\mathcal X_{2^\sigma}\cdots\mathcal X_{N^\sigma}\rangle$. Cyclic and reflection symmetries of the traces imply that
\begin{equation}\label{eq:dihedral}
    G^{\mathcal X}_{12\cdots N}=G^{\mathcal X}_{2\cdots N1}=(-)^NG^{\mathcal X}_{N\cdots21}.
\end{equation}
We focus on $G^{\mathcal X}_{12\cdots N}$ since any other color ordering can then be obtained by relabeling. 

At tree level, the correlator $\mathcal G^{\mathcal O}_{12\cdots N}$ of any multiplicity $N$ can be constructed recursively \cite{Cao:2023cwa,Cao:2024bky} with explicit results up to 8 points. The recursion generates all single-gluon amplitudes $\langle\mathcal O_1\cdots\mathcal O_{N-1}\mathcal J_N\rangle$ as a byproduct. However, there are no explicit results for the complete amplitude involving two or more gluons. Even for the four-point function, where one could in principle use superspace methods to compute correlators such as $\langle\mathcal J\mathcal J\mathcal O\mathcal O\rangle$ and $\langle\mathcal J\mathcal J\mathcal J\mathcal J\rangle$ from $\langle\mathcal O\mathcal O\mathcal O\mathcal O\rangle$, the currently available results \cite{Bissi:2022wuh} are limited to the so-called ``orthogonal frame'', where $Z_p\cdot X_q=0$, due to the complicated polarization structure. Below, we remedy this by bootstrapping the complete results of $\langle\mathcal J\mathcal J\mathcal O\mathcal O\rangle$ and $\langle\mathcal J\mathcal J\mathcal J\mathcal J\rangle$ with a simple calculation, which matches their results when reduced to the orthogonal frame.

Let us first recall the important consequences of color-ordering. First, it induces a natural set of planar Mandelstam variables in the amplitude. The momentum flowing through a propagator in the amplitude can only be a sum of consecutive external momenta. This leads us to introduce planar Mellin variables for the total momentum between the $p$-th and $(q-1)$-th operators:
\begin{equation}\label{eq:chidef}
    \chi_{p,q}:=\sum_{\substack{p\leq r < q\\ q\leq s <p}}(\gamma_{rs}-\eta_{r's}-\eta_{rs'}-\eta_{r's'}) = -\left( \sum_{p\leq r < q} (\mathtt k_r + \mathtt k_{r'}) \right)^2.
\end{equation}
Here, all the ranges of labels should be understood as modulo $N$. For example, the range $p\leq r < q$ means 
$r\in \{ p,p+1,\cdots,q-2,q-1 \}$ when $p<q$, and $r\in \{ p,p+1,\cdots,N,1,\cdots ,q-2,q-1 \}$ when $p>q$. Color-ordered amplitudes only have poles of the form $\frac1{\chi_{p,q}-\tau-2m}$. In total, we have $\frac{N(N-3)}2$ planar variables, which can be used to solve for all continuous Mellin variables $\gamma_{pq}$:
\begin{equation}
    \gamma_{pq}=\eta_{p'q}+\eta_{pq'}+\eta_{p'q'}+\frac{\chi_{p,q}-\chi_{p,q+1}-\chi_{p+1,q}+\chi_{p+1,q+1}}2,
\end{equation}
where we define $\chi_{p,p+1}:=\tau_p$ and $\chi_{p,p}:=0$. The color-ordering also induces a nice basis of R-structures. Denoting $\langle pq\rangle:=v_p^\alpha v_q^\beta\epsilon_{\alpha\beta}$, the color-ordered amplitudes can be expanded onto products of non-crossing cycles $V_{p_1p_2\cdots p_n}:=\langle p_1p_2\rangle\langle p_2p_3\rangle\cdots\langle p_n p_1\rangle$, where each $v_p^\alpha$ represents the R-polarization of the $p$-th operator.

In the $\mathcal N=2$ supergluon theory we are interested in, both $\mathcal O$ and $\mathcal J$ have $\tau=2$. Therefore, these two operators contribute to the same series of poles $\frac{1}{\chi-2-2m}$ in the factorization and their contributions mix together in the residues. However, since the two operators have different $SU(2)_R$ charges, we can separate their individual contributions by projecting out the R-spin-1 and R-spin-0 parts~\cite{Cao:2023cwa,Cao:2024bky}. We can thus define the gluon residue
\begin{equation}\label{eq:extractGluon}
    \mathop{\rm Res}_{\chi_{p,q}=\tau+2m}^{(\mathcal J)}\mathcal M=\left.\mathop{\rm Res}_{\chi_{p,q}=\tau+2m}\mathcal M\right|_{V_{i\cdots jk\cdots l}\mapsto\frac12V_{i\cdots j}V_{k\cdots l}},\quad\forall p\leq i<j<q,\ q\leq k<l<p,
\end{equation}
which extracts the contribution of $\mathcal J$ in the channel $\frac{1}{\chi_{p,q}-\tau-2m}$, and define the scalar residue 
\begin{equation}\label{eq:extractScalar}
    \mathop{\rm Res}_{\chi_{p,q}=\tau+2m}^{(\mathcal O)}\mathcal M= \mathop{\rm Res}_{\chi_{p,q}=\tau+2m}\mathcal M - \mathop{\rm Res}_{\chi_{p,q}=\tau+2m}^{(\mathcal J)}\mathcal M
\end{equation}
by subtracting the gluon contributions. As an example, from the residue of the four-point amplitude
\begin{equation}
    \mathop{\rm Res}_{\chi_{1,3}=2}\mathcal M^{\mathcal O}_{1234}=-2V_{1234}-V_{12}V_{34}(\chi_{2,4}-4),
\end{equation}
we can read off the gluon and scalar contributions:
\begin{equation}
    \mathop{\rm Res}_{\chi_{1,3}=2}^{(\mathcal J)}\mathcal M^{\mathcal O}_{1234}=-V_{12}V_{34}(\chi_{2,4}-3),\quad \mathop{\rm Res}_{\chi_{1,3}=2}^{(\mathcal O)}\mathcal M^{\mathcal O}_{1234}=-2\left(V_{1234}-\frac12V_{12}V_{34}\right).
\end{equation}

The Mellin amplitudes for $\langle\mathcal O_1\cdots\mathcal O_N\rangle$ ($N\leq8$) and $\langle\mathcal O_1\cdots\mathcal O_{N-1}\mathcal J_N\rangle$ ($N\leq6$) can be obtained from \href{https://journals.aps.org/prl/supplemental/10.1103/PhysRevLett.133.021605/results_sca8_glu6.txt}{\texttt{this URL}}. The conventions of the data file were described in detail in~\cite{Cao:2023cwa}\footnote{Be sure to use the published version or the latest arXiv version (v4), as previous preprint versions contain typos in the ancillary file.}. In our formalism,
\begin{equation}\label{eq:3pt}
    \mathcal M^{\mathcal O}_{123}=V_{123},\quad\mathcal M_{\mathcal O_1\mathcal O_2\mathcal J_3}=\frac{i}{2\sqrt3}V_{12}(\eta_{13'}-\eta_{23'}). 
\end{equation}
The planar variables in the ancillary files are related to the planar variables defined in this paper through:
\begin{equation}
    \chi_{p,q}=2-\texttt{xx[$p$,$q$]}.
\end{equation}

\subsubsection{$\langle\mathcal J\mathcal J\mathcal J\rangle$}

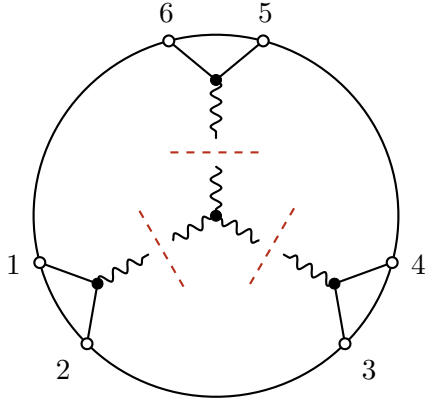
\begin{figure}
    \centering
    \begin{tikzpicture}[scale=1.2]
        \fill [white] (2.3,2.3) rectangle (-2.3,-2.3);
        \draw [thick] (0,0) circle [radius=2];
        \draw [fill] (0,0) circle [radius=0.06];
        \foreach \angle in {0,120,240}{
            \begin{scope}[rotate=\angle]
              \draw [thick] (90-15:2) -- (90:1.5) -- (90+15:2); 
              \draw [thick,decorate,decoration={snake, amplitude=2, segment length=6.28}]  (90:1.5) -- (90:0.85);
              \draw [thick,decorate,decoration={snake, amplitude=2, segment length=6.28}]  (90:0.55) -- (0,0);
              \draw [fill] (90:1.5) circle [radius=0.06];
              \draw [thick, fill=white] (90-15:2) circle [radius=0.06];
              \draw [thick, fill=white] (90+15:2) circle [radius=0.06];
              \draw [thick,dashed,BrickRed] (-0.5,0.7) -- (0.5,0.7);
            \end{scope}
        };
        \node [anchor = east] at (210-15:2.1) {$1$};
        \node [anchor = north east] at (210+15:2.1) {$2$};
        \node [anchor = north west] at (330-15:2.1) {$3$};
        \node [anchor = west] at (330+15:2.1) {$4$};
        \node [anchor = south] at (90-15:2.1) {$5$};
        \node [anchor = south] at (90+15:2.1) {$6$};
    \end{tikzpicture}
    \caption{The factorization of $\mathcal M^{\mathcal O}_{123456}$ on gluons in the snowflake channel. The residue is related to the product of $\mathcal{M}_{\mathcal J \mathcal J \mathcal J} $ and three $\mathcal M_{\mathcal O \mathcal O \mathcal J}$.}
    \label{fig:snowflake}
\end{figure}

As discussed in the previous subsection, gauge-invariance and the unitarity bound dictate that the three-point gluon Mellin amplitude $\mathcal M^{\mathcal J}_{123}$ must be a linear combination of two structures. Plugging in $\Delta=3$, we have
\begin{equation}
    \begin{aligned}
        \mathcal M^{\mathcal J}_{123}&=c\Big(\eta_{1'2'}(\eta_{3'1}-\eta_{3'2})+\eta_{2'3'}(\eta_{1'2}-\eta_{1'3})+\eta_{3'1'}(\eta_{2'3}-\eta_{2'1})\\
        &\qquad+(\eta_{1'2}\eta_{2'3}\eta_{3'1}-\eta_{1'3}\eta_{2'1}\eta_{3'2})\Big)\\
        &-c'\Big((\eta_{1'2}-\eta_{1'3})(\eta_{2'3}-\eta_{2'1})(\eta_{3'1}-\eta_{3'2})\\
        &\qquad+(\eta_{1'2}\eta_{2'3}\eta_{3'1}-\eta_{1'3}\eta_{2'1}\eta_{3'2})\Big).
    \end{aligned}
\end{equation}
To determine the values of $c,c'$ in this theory, we can glue it with three $\mathcal M_{\mathcal O\mathcal O\mathcal J}$ amplitudes in \eqref{eq:3pt}, and compare the result with the contribution from three gluon exchanges in the snowflake channel in $\mathcal M^{\mathcal O}_{123456}$, as is shown in figure \ref{fig:snowflake}. 

First, using the factorization formula derived in Section~\ref{sec:factorization}, we can glue $\mathcal M^{\mathcal J}_{12L}\otimes\mathcal M_{\mathcal O_3\mathcal O_4\mathcal J_{R}}$ to obtain\footnote{We have simplified the result a bit by dropping terms of the form $\eta^{\underline2}$ because both operators 1 and 2 have spin-1.}
\begin{align*}
    &\left(i\sqrt3V_{34}\right)^{-1}\mathop{\rm Res}^{(\mathcal J)}_{\chi_{1,3}=2}\mathcal M_{\mathcal J_1\mathcal J_2\mathcal O_3\mathcal O_4}=\left(i\sqrt3V_{34}\right)^{-1}\left.\mathcal K_{3,1}\hat{\mathbb N}(\mathcal M^{\mathcal J}_{12L}\mathcal M_{\mathcal O_3\mathcal O_4\mathcal J_{R}})\right|_{\substack{\eta_{aL'}=0\\\eta_{iR'}=0}}\\
    =\,&c\Big(\chi_{2,4}(2\eta_{1'2'}+\eta_{1'2}\eta_{2'\{34\}}+\eta_{2'1}\eta_{1'\{34\}})\\
    &\quad{-}2(3\eta_{1'2'}+2(\eta_{1'2}\eta_{2'3}+\eta_{2'1}\eta_{1'4})+\eta_{1'2}\eta_{2'4}+\eta_{2'1}\eta_{1'3}+\eta_{1'3}\eta_{2'4}-\eta_{1'4}\eta_{2'3})\Big)\\
    +\,&c'\Big(\chi_{2,4}(2\eta_{1'2}\eta_{2'1}+2\eta_{1'\{34\}}\eta_{2'\{34\}}-3\eta_{1'2}\eta_{2'\{34\}}-3\eta_{2'1}\eta_{1'\{34\}})\\
    &\quad{-}2(3\eta_{1'2}\eta_{2'1}+2\eta_{1'3}\eta_{2'4}+3\eta_{1'3}\eta_{2'3}+3\eta_{1'4}\eta_{2'4}\\
    &\quad\quad\quad{+}4\eta_{1'4}\eta_{2'3}-4\eta_{1'2}\eta_{2'4}-4\eta_{2'1}\eta_{1'3}-5\eta_{1'2}\eta_{2'3}-5\eta_{2'1}\eta_{1'4})\Big).
\end{align*}
Here, we have used short-hand notations such as $\eta_{1'\{34\}}:=\eta_{1'3}+\eta_{1'4}$. Gluing another two copies of $\mathcal M_{\mathcal O\mathcal O\mathcal J}$, we obtain
\begin{align*}
    &\left(-6i\sqrt3V_{12}V_{34}V_{56}\right)^{-1}\mathop{\rm Res}^{(\mathcal J)}_{\chi_{1,3}=2}\mathop{\rm Res}^{(\mathcal J)}_{\chi_{3,5}=2}\mathop{\rm Res}^{(\mathcal J)}_{\chi_{1,5}=2}\mathcal M^{\mathcal O}_{123456}\\
    =\,&c\Big(\chi_{1,4}\chi_{2,5}\chi_{3,6}-36+4(\chi_{1,4}+\chi_{2,5}+\chi_{3,6})+6(\chi_{2,4}+\chi_{4,6}+\chi_{2,6})\\
    &\quad{-}(\chi_{1,4}\chi_{2,5}+\chi_{2,5}\chi_{3,6}+\chi_{3,6}\chi_{1,4})-2(\chi_{1,4}\chi_{2,6}+\chi_{2,5}\chi_{4,6}+\chi_{3,6}\chi_{2,4})\Big)\\
    +\,&c'\Big(144-5\chi_{1,4}\chi_{2,5}\chi_{3,6}-40(\chi_{1,4}+\chi_{2,5}+\chi_{3,6})-6(\chi_{2,4}+\chi_{4,6}+\chi_{2,6})\\
    &\quad{+}13(\chi_{1,4}\chi_{2,5}+\chi_{2,5}\chi_{3,6}+\chi_{3,6}\chi_{1,4})+2(\chi_{1,4}\chi_{2,6}+\chi_{2,5}\chi_{4,6}+\chi_{3,6}\chi_{2,4})\Big).
\end{align*}
The correct result can be extracted from $\mathcal M^{\mathcal O}_{123456}$ using \eqref{eq:extractGluon}:
\begin{align*}
    &\left(2V_{12}V_{34}V_{56}\right)^{-1}\mathop{\rm Res}^{(\mathcal J)}_{\chi_{1,3}=2}\mathop{\rm Res}^{(\mathcal J)}_{\chi_{3,5}=2}\mathop{\rm Res}^{(\mathcal J)}_{\chi_{1,5}=2}\mathcal M^{\mathcal O}_{123456}\\
    =\,&6(\chi_{2,4}+\chi_{4,6}+\chi_{2,6})-5(\chi_{1,4}+\chi_{2,5}+\chi_{3,6})-9\\
    +\,&2(\chi_{1,4}\chi_{2,5}+\chi_{2,5}\chi_{3,6}+\chi_{3,6}\chi_{1,4})-2(\chi_{1,4}\chi_{2,6}+\chi_{2,5}\chi_{4,6}+\chi_{3,6}\chi_{2,4}).
\end{align*}
Comparing the two results, we fix the unknown coupling constants in this theory to be
\begin{equation}
    c=\frac{5i}{12\sqrt3},\quad c'=\frac{i}{12\sqrt3}.
\end{equation}
In particular, since in pure Yang-Mills theory, $c'/c=1/(2d-3)=1/5$ in $d=4$ (see the comments in subsection~\ref{subsec:gluon3gen}), we have verified that the effective bulk Lagrangian of this $\mathcal N=2$ theory does not contain an $F^3$ interaction term.

\subsubsection{$\langle\mathcal J\mathcal J\mathcal J\mathcal J\rangle$} 

\begin{figure}
    \centering
    \begin{tikzpicture}[scale=0.7]
    \begin{scope}
        \draw [thick] (0,0) circle [radius=2];
        \foreach \angle in {0,90,180,270}{
            \begin{scope}[rotate=\angle]
              \draw [thick,decorate,decoration={snake, amplitude=2, segment length=6.28}]  (135:2) -- (135:0);
              \draw [thick, fill=white] (135:2) circle [radius=0.08];
            \end{scope}
        };
        \draw [thick, fill=gray!50] (0,0) circle [radius=0.8];
        \node [anchor = south east] at (135:2.1) {$1$};
        \node [anchor = north east] at (-135:2.1) {$2$};
        \node [anchor = north west] at (-45:2.1) {$3$};
        \node [anchor = south west] at (45:2.1) {$4$};
    \end{scope}
    \begin{scope}[xshift=3cm]
        \node at (0,0) {$=$};
    \end{scope}
    \begin{scope}[xshift=6cm]
        \draw [thick] (0,0) circle [radius=2];
        \draw [thick,decorate,decoration={snake, amplitude=2, segment length=6.28}]  (-135:2) -- (180:0.8) -- (135:2);
        \draw [thick,decorate,decoration={snake, amplitude=2, segment length=6.28}]  (45:2) -- (0:0.8) -- (-45:2);
        \draw [thick,decorate,decoration={snake, amplitude=2, segment length=6.28}] (0:0.8) -- (180:0.8);
        \draw [fill] (180:0.8) circle [radius=0.08];
        \draw [fill] (0:0.8) circle [radius=0.08];
        \foreach \angle in {0,90,180,270}{
            \begin{scope}[rotate=\angle]
              \draw [thick, fill=white] (135:2) circle [radius=0.08];
            \end{scope}
        };
        \node [anchor = south east] at (135:2.1) {$1$};
        \node [anchor = north east] at (-135:2.1) {$2$};
        \node [anchor = north west] at (-45:2.1) {$3$};
        \node [anchor = south west] at (45:2.1) {$4$};
    \end{scope}
    \begin{scope}[xshift=9cm]
        \node at (0,0) {$+$};
    \end{scope}
    \begin{scope}[xshift=12cm]
        \begin{scope}[rotate=90]
            \draw [thick] (0,0) circle [radius=2];
            \draw [thick,decorate,decoration={snake, amplitude=2, segment length=6.28}]  (-135:2) -- (180:0.8) -- (135:2);
            \draw [thick,decorate,decoration={snake, amplitude=2, segment length=6.28}]  (45:2) -- (0:0.8) -- (-45:2);
            \draw [thick,decorate,decoration={snake, amplitude=2, segment length=6.28}] (0:0.8) -- (180:0.8);
            \draw [fill] (180:0.8) circle [radius=0.08];
            \draw [fill] (0:0.8) circle [radius=0.08];
            \foreach \angle in {0,90,180,270}{
                \begin{scope}[rotate=\angle]
                  \draw [thick, fill=white] (135:2) circle [radius=0.08];
                \end{scope}
            };
        \end{scope}
        \node [anchor = south east] at (135:2.1) {$1$};
        \node [anchor = north east] at (-135:2.1) {$2$};
        \node [anchor = north west] at (-45:2.1) {$3$};
        \node [anchor = south west] at (45:2.1) {$4$};
    \end{scope}
    \begin{scope}[xshift=15cm]
        \node at (0,0) {$+$};
    \end{scope}
    \begin{scope}[xshift=18cm]
        \draw [thick] (0,0) circle [radius=2];
        \foreach \angle in {0,90,180,270}{
            \begin{scope}[rotate=\angle]
              \draw [thick,decorate,decoration={snake, amplitude=2, segment length=6.28}]  (135:2) -- (135:0);
              \draw [thick, fill=white] (135:2) circle [radius=0.08];
            \end{scope}
        };
        \draw [fill] (0,0) circle [radius=0.08];
        \node [anchor = south east] at (135:2.1) {$1$};
        \node [anchor = north east] at (-135:2.1) {$2$};
        \node [anchor = north west] at (-45:2.1) {$3$};
        \node [anchor = south west] at (45:2.1) {$4$};
    \end{scope}
    \end{tikzpicture}
    \caption{Possible diagrams in the four-gluon amplitude $\mathcal M^{\mathcal J}_{1234}$.}
    \label{fig:JJJJ}
\end{figure}
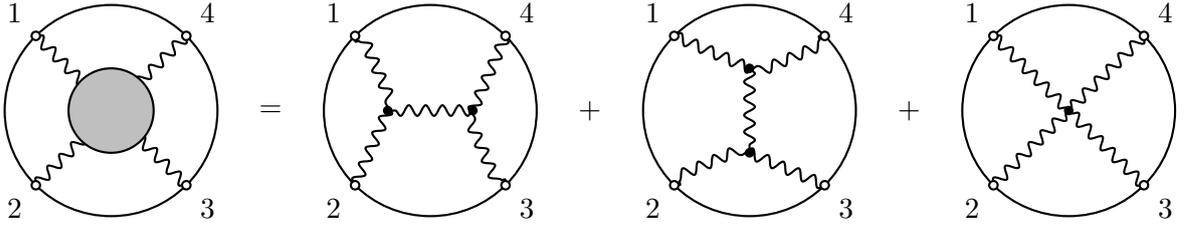

With the three-gluon amplitude $\mathcal M^{\mathcal J}_{123}$ fixed, we can now proceed to bootstrap the four-gluon amplitude $\mathcal M^{\mathcal J}_{1234}$. Note that both exchange diagrams and contact diagrams contribute to this amplitude, as illustrated in figure \ref{fig:JJJJ}.

For the exchange diagrams, we can simply glue two three-gluon amplitudes together. Note that we have to take into account all descendant contributions, namely,
\begin{equation}
    \begin{aligned}
        (\mathcal M^{\mathcal J}_{1234})_{\text{exchange}}&=\left.\sum_{m=0}^\infty\frac{\frac{\mathcal K_{3,1}}{4^mm!(m+1)!}\hat{\mathbb N}\Big[(\hat{\mathbbm x}^m\mathcal M^{\mathcal J}_{12 R})(\hat{\mathbbm x}^m\mathcal M^{\mathcal J}_{ L34})\Big]}{\chi_{1,3}-2-2m}\right|_{\eta_{aL'}=\eta_{iR'}=0}\\
        &+\left.\sum_{m=0}^\infty\frac{\frac{\mathcal K_{3,1}}{4^mm!(m+1)!}\hat{\mathbb N}\Big[(\hat{\mathbbm x}^m\mathcal M^{\mathcal J}_{23 R})(\hat{\mathbbm x}^m\mathcal M^{\mathcal J}_{ L41})\Big]}{\chi_{2,4}-2-2m}\right|_{\eta_{aL'}=\eta_{iR'}=0}.
    \end{aligned}
\end{equation}
However, it can be easily checked that $\hat{\mathbbm x}^2\mathcal M^{\mathcal J}_{12 R}=0$, which implies that the descendant poles truncate to only $m=0,1$. The result is too lengthy to write down explicitly. Here, we merely point out that the result of $(\mathcal M^{\mathcal J}_{1234})_{\text{exchange}}$ is already gauge-invariant up to contact terms:
\begin{equation*}
    \hat{\mathbbm m}(\mathcal M^{\mathcal J}_{123 R})_{\text{exchange}}=\text{contact terms, i.e., no poles of the form }\frac1{\chi-\#}.
\end{equation*}
This suggests that we should use gauge-invariance to constrain the contact terms.

To do this, we write down an ansatz for $(\mathcal M^{\mathcal J}_{1234})_{\text{contact}}$ that is a polynomial\footnote{In principle, we could also write a larger ansatz that also depends on $\gamma$'s. However, since $(\mathcal M^{\mathcal J}_{1234})_{\text{exchange}}\sim\chi^0$ as $\chi\to\infty$, we expect $(\mathcal M^{\mathcal J}_{1234})_{\text{contact}}$ to also obey this power-counting, which restricts it to be a polynomial of $\eta$'s only. This will hopefully be justified by a careful analysis of the flat-space limit of spinning Mellin amplitudes, which we leave for future work.} of $\eta$'s and multi-linear in the labels $1',2',3',4'$. The ansatz has 138 unknown coefficients, and dihedral symmetry\footnote{The exchange contributions automatically satisfy the dihedral symmetry, as can be checked explicitly.}~\eqref{eq:dihedral} reduces the number of unknowns to 27. We then impose the gauge-invariance condition, which reads
\begin{equation}
    \hat{\mathbbm m}(\mathcal M^{\mathcal J}_{123 R})_{\text{exchange}}+\hat{\mathbbm m}(\mathcal M^{\mathcal J}_{123 R})_{\text{contact}}=0,\quad\text{on lattice points}.
\end{equation}
Remarkably, this uniquely fixes all unknown coefficients! The complete result of $\mathcal M^{\mathcal J}_{1234}$ is recorded in an ancillary file, which agrees with the snow-flake channel of $\mathcal M^{\mathcal O}_{12345678}$ when glued with four copies of $\mathcal M_{\mathcal O\mathcal O\mathcal J}$.

The result of $\mathcal M^{\mathcal J}_{1234}$ in the orthogonal frame has been obtained in (8.21) in \cite{Bissi:2022wuh}. The orthogonal frame is the special configuration $Z_p\cdot X_q=0$ of the correlator, which is equivalent to taking all $\eta_{p'q}=0$ in the spinning Mellin amplitudes. In our notation, their result reads:
\begin{align}
    \langle\mathcal J\mathcal J\mathcal J\mathcal J\rangle_{\rm ortho}&=Z_{12}Z_{34}\int{\rm d}\gamma\,\widetilde{\mathcal M}^{(1)}(\gamma)\prod_{1\leq p<q\leq4}\frac{\Gamma(\gamma_{pq})}{(-X_{pq})^{\gamma_{pq}}}\nonumber\\
    &+Z_{14}Z_{23}\int{\rm d}\gamma\,\widetilde{\mathcal M}^{(2)}(\gamma)\prod_{1\leq p<q\leq4}\frac{\Gamma(\gamma_{pq})}{(-X_{pq})^{\gamma_{pq}}}\nonumber\\
    &+Z_{13}Z_{24}\int{\rm d}\gamma\,\widetilde{\mathcal M}^{(3)}(\gamma)\prod_{1\leq p<q\leq4}\frac{\Gamma(\gamma_{pq})}{(-X_{pq})^{\gamma_{pq}}}.
\end{align}
These Mellin variables satisfy
\begin{equation}
    \gamma_{12}+\gamma_{13}+\gamma_{14}=\Delta_1=3,\quad\text{and cyclic},
\end{equation}
which relate to the Mellin variables $\mathrm s,\mathrm t,\mathrm u$ in \cite{Bissi:2022wuh} by
\begin{equation}
    \mathrm s=\Delta_1+\Delta_2-2\gamma_{12}=6-2\gamma_{12},\quad\mathrm t=6-2\gamma_{23},\quad\mathrm u=6-2\gamma_{13},\quad\mathrm s+\mathrm t+\mathrm u=12.
\end{equation}
On the support of $\eta_{p'q}=0$, these are related to our $\chi$ variables through: 
\begin{subequations}
\begin{gather}
    \chi_{1,3}=\tau_1+\tau_2-2\gamma_{12}+2\eta_{1'2'}=\mathrm s+2\eta_{1'2'}-2=\begin{cases}
    \mathrm s,&\text{in }\widetilde{\mathcal M}^{(1)},\\
    \mathrm s-2,&\text{in }\widetilde{\mathcal M}^{(2,3)},
    \end{cases}\\
    \chi_{2,4}=\tau_2+\tau_3-2\gamma_{23}+2\eta_{2'3'}=\mathrm t+2\eta_{2'3'}-2=\begin{cases}
    \mathrm t,&\text{in }\widetilde{\mathcal M}^{(2)},\\
    \mathrm t-2,&\text{in }\widetilde{\mathcal M}^{(1,3)}.
    \end{cases}
\end{gather}
\end{subequations}
Up to an overall factor, the Mellin amplitude in~\cite{Bissi:2022wuh} reads\footnote{Note that the coefficient of $\mathsf c_t-\mathsf c_u$ in $\widetilde{\mathcal M}^{(1)}$ in that paper contains a typo. It should be $-\frac{33}2$ instead of $-33$.}
\begin{equation}
\begin{gathered}
    \widetilde{\mathcal M}^{(1)}=\mathsf c_s\frac{25(\mathrm t-\mathrm u)}{2(\mathrm s-2)}+\mathsf c_s\frac{4(\mathrm t-\mathrm u)}{\mathrm s-4}-\mathsf c_t\frac{\mathrm s+21}{\mathrm t-4}+\mathsf c_u\frac{\mathrm s+21}{\mathrm u-4}-\frac{33}2(\mathsf c_t-\mathsf c_u),\\[1mm]
    \widetilde{\mathcal M}^{(2)}=-\widetilde{\mathcal M}^{(1)}|_{\substack{\mathrm s\leftrightarrow\mathrm t\\\mathsf c_s\leftrightarrow\mathsf c_t}},\quad\widetilde{\mathcal M}^{(3)}=-\widetilde{\mathcal M}^{(1)}|_{\substack{\mathrm s\leftrightarrow\mathrm u\\\mathsf c_s\leftrightarrow\mathsf c_u}}.
\end{gathered}
\end{equation}
Here, the color factors are related to traces through
\begin{subequations}
    \begin{gather}
        \begin{aligned}
        \mathsf c_s= f^{a_1a_2b}f^{b\,a_3a_4}\propto&\ \mathop{\rm tr}(T^{a_1}T^{a_2}T^{a_3}T^{a_4})-\mathop{\rm tr}(T^{a_1}T^{a_2}T^{a_4}T^{a_3}) \\
        &\quad -\mathop{\rm tr}(T^{a_1}T^{a_3}T^{a_4}T^{a_2})+\mathop{\rm tr}(T^{a_1}T^{a_4}T^{a_3}T^{a_2}),
        \end{aligned}\\[1mm]
        \mathsf c_t=f^{a_1a_4b}f^{b\,a_2a_3}= -\mathsf c_s \big|_{a_2 \leftrightarrow a_4},\qquad \mathsf c_u=f^{a_1a_3b}f^{b\,a_4a_2}= -\mathsf c_s |_{a_2 \leftrightarrow a_3}.
    \end{gather}
\end{subequations}
To extract the partial amplitude we are considering, one can simply take 
\begin{equation}
    \mathsf c_s \mapsto 1,\qquad  \mathsf c_t \mapsto -1,\qquad \mathsf c_u \mapsto 0,
\end{equation}
which gives the coefficient of $\mathop{\rm tr}(T^{a_1}T^{a_2}T^{a_3}T^{a_4})$ (up to an overall coefficient). With these dictionaries, we can easily check that our results agree with those of \cite{Bissi:2022wuh} up to an overall normalization:
\begin{equation}
\begin{aligned}
    -36\mathcal M^{\mathcal J}_{1234}\Big|_{\eta_{1'2'}=\eta_{3'4'}=1}&=\frac{25(\chi_{2,4}-3)}{\chi_{1,3}-2}+\frac{8(\chi_{2,4}-2)}{\chi_{1,3}-4}+\frac{\chi_{1,3}+21}{\chi_{2,4}-2}+33\\
    &=\frac{25(\mathrm t-5)}{\mathrm s-2}+\frac{8(\mathrm t-4)}{\mathrm s-4}+\frac{\mathrm s+21}{\mathrm t-4}+33\\
    &=\widetilde{\mathcal M}^{(1)}\Big|_{\substack{\mathrm u=12-\mathrm s-\mathrm t\\\mathsf c_s=+1,\mathsf c_t=-1,\mathsf c_u=0}},
\end{aligned}
\end{equation}
and
\begin{align}
    -36\mathcal M^{\mathcal J}_{1234}\Big|_{\eta_{1'4'}=\eta_{2'3'}=1}&=\widetilde{\mathcal M}^{(2)}\Big|_{\substack{\mathrm u=12-\mathrm s-\mathrm t\\\mathsf c_s=+1,\mathsf c_t=-1,\mathsf c_u=0}},\\
-36\mathcal M^{\mathcal J}_{1234}\Big|_{\eta_{1'3'}=\eta_{2'4'}=1}&=\widetilde{\mathcal M}^{(3)}\Big|_{\substack{\mathrm u=12-\mathrm s-\mathrm t\\\mathsf c_s=+1,\mathsf c_t=-1,\mathsf c_u=0}},
\end{align}
as can be checked.

\subsubsection{$\langle\mathcal J\mathcal J\mathcal O\mathcal O\rangle$}

We have also computed\footnote{Due to color-ordering, there are two inequivalent color-ordered amplitudes that we need to compute.} $\langle\mathcal J_1\mathcal J_2\mathcal O_3\mathcal O_4\rangle$ and $\langle\mathcal J_1\mathcal O_2\mathcal J_3\mathcal O_4\rangle$ using the same strategy. Since $\hat{\mathbbm x}\mathcal M_{\mathcal O_1\mathcal O_2\mathcal J_{ R}}=0$ and $\hat{\mathbbm x}^2\mathcal M_{\mathcal O_1\mathcal O_{ R}\mathcal J_3}=0$, the exchange contributions have truncated poles:
\begin{subequations}
\begin{align}
    &(\mathcal M_{\mathcal J_1\mathcal J_2\mathcal O_3\mathcal O_4})_{\text{exchange}}\nonumber\\
    =\,&\left.\mathcal K_{2,0}\frac{\mathcal M_{\mathcal O_4\mathcal J_1\mathcal O_{ R}}\mathcal M_{\mathcal J_2\mathcal O_3\mathcal O_{ L}}}{\chi_{2,4}-2}+\frac{\mathcal K_{2,0}}4\frac{(\hat{\mathbbm x}\mathcal M_{\mathcal O_4\mathcal J_1\mathcal O_{ R}})(\hat{\mathbbm x}\mathcal M_{\mathcal J_2\mathcal O_3\mathcal O_{ L}})}{\chi_{2,4}-4}\right|_{\eta_{aL'}=\eta_{iR'}=0}\nonumber\\
    +\,&\left.\mathcal K_{3,1}\frac{\hat{\mathbb N}\Big[\mathcal M^{\mathcal J}_{12 R}\mathcal M_{\mathcal O_3\mathcal O_4\mathcal J_{ L}}\Big]}{\chi_{1,3}-2}\right|_{\eta_{aL'}=\eta_{iR'}=0},\\
    &(\mathcal M_{\mathcal J_1\mathcal O_2\mathcal J_3\mathcal O_4})_{\text{exchange}}\nonumber\\
    =\,&\left.\mathcal K_{2,0}\frac{\mathcal M_{\mathcal O_4\mathcal J_1\mathcal O_{ R}}\mathcal M_{\mathcal O_{ L}\mathcal O_2\mathcal J_3}}{\chi_{2,4}-2}+\frac{\mathcal K_{2,0}}4\frac{(\hat{\mathbbm x}\mathcal M_{\mathcal O_4\mathcal J_1\mathcal O_{ R}})(\hat{\mathbbm x}\mathcal M_{\mathcal O_{ L}\mathcal O_2\mathcal J_3})}{\chi_{2,4}-4}\right|_{\eta_{aL'}=\eta_{iR'}=0}\nonumber\\
    +\,&\left.\mathcal K_{2,0}\frac{\mathcal M_{\mathcal J_1\mathcal O_2\mathcal O_{ R}}\mathcal M_{\mathcal J_3\mathcal O_4\mathcal O_{ L}}}{\chi_{1,3}-2}+\frac{\mathcal K_{2,0}}4\frac{(\hat{\mathbbm x}\mathcal M_{\mathcal J_1\mathcal O_2\mathcal O_{ R}})(\hat{\mathbbm x}\mathcal M_{\mathcal J_3\mathcal O_4\mathcal O_{ L}})}{\chi_{1,3}-4}\right|_{\eta_{aL'}=\eta_{iR'}=0}.
\end{align}
\end{subequations}
Again, it can be checked that the exchange contribution is already gauge-invariant up to contact terms. The ansatz for the contact contribution can be uniquely fixed by gauge-invariance and dihedral symmetry, and the final results are recorded in an ancillary file. The orthogonal frame result can be extracted by setting $\eta_{p'q}=0$, which agrees with (8.11) of \cite{Bissi:2022wuh} up to an overall normalization.

\section{Summary and outlook}\label{sec:outlook}

In this paper, we have proposed a definition of spinning Mellin amplitudes for any number of external spinning operators. Compared with previous attempts in the literature, our definition is easy to use and particularly suitable for the computation/bootstrap of higher-point spinning correlators. Moreover, our definition highlights the similarity between a spinning operator and multiple (two, in the case of symmetric traceless tensors) scalar operators, which generalizes the factorization property of scalar Mellin amplitudes to spinning Mellin amplitudes in a straightforward manner. Following the strategy proposed in \cite{Goncalves:2014rfa} of solving Casimir recurrence relations, we have developed a systematic algorithm to obtain the factorization formula for the exchange of spinning operators at any descendant level. We have demonstrated the power of our formalism by bootstrapping three- and four-point gluon amplitudes, with complete polarization dependence, in the model of supergluons in $\mathrm{AdS}_5 \times \mathrm{S}^3$. Now that we have a framework to deal with spinning correlators, several natural questions can be addressed.

First, it will be very important to investigate the structure and properties of spinning Mellin amplitudes in more detail. Our bootstrap calculations in section~\ref{sec:boots} suggest that there could be a direct relation between the effective bulk Lagrangian and spinning Mellin amplitudes. It will be very interesting to clarify this relation, or even to find a set of Feynman rules in Mellin space, similar to the scalar Feynman rules in~\cite{Paulos:2011ie}. It is also interesting to explore the question of uniqueness in Mellin space: are certain three-point spinning correlators uniquely characterized by certain fundamental principles such as gauge-invariance or unitarity, similar to flat-space amplitudes~\cite{Arkani-Hamed:2016rak}? Previous studies in position space include \cite{Stanev:2012nq,Elkhidir:2014woa,Buchbinder:2022cqp,Buchbinder:2023coi} in specific dimensions and \cite{Osborn:1993cr,Erdmenger:1996yc,Zhiboedov:2012bm,Karapetyan:2023zdu,Karapetyan:2025ick} in general dimensions. The Mellin formalism may provide a natural language to discuss this problem further. 

In addition, a careful analysis of contact Witten diagrams along the lines of~\cite{Penedones:2010ue,Paulos:2011ie} could elucidate the correct prescription of taking flat-space limits of spinning Mellin amplitudes. In particular, it will be highly non-trivial to see the appearance of gauge invariance $\mathtt e\sim\mathtt e+\mathtt k$ in flat space. The fact that $N$-point spinning Mellin amplitudes look like $2N$-point scalar Mellin amplitudes reminds us of the recently introduced ``scaffolding'' perspective on flat-space spinning amplitudes~\cite{Arkani-Hamed:2023jry}, which could play an important role in understanding the flat-space limit of spinning Mellin amplitudes. Hopefully, the flat-space limit will also be understood physically, as in \cite{Fitzpatrick:2011hu,Paulos:2016fap} (also see \cite{Li:2021snj} for a literature review on the flat-space limit in other representations). In practice, a good knowledge of the flat-space limit dictates the correct power-counting of Mellin amplitudes, which will be very helpful in the bootstrap of higher-point correlators.

We are optimistic that our spinning Mellin amplitudes will facilitate the bootstrap of higher-point correlators and the study of their properties in specific theories. On one hand, sub-amplitudes with multiple spinning operators can now be computed independently, which can be glued together to obtain high-multiplicity amplitudes. On the other hand, we are finally in a position to explore the structure of higher-point spinning Mellin amplitudes. For example, it may be possible to reconcile the BCJ relations observed for spinning correlators in differential representations~\cite{Diwakar:2021juk,Herderschee:2022ntr} and the nonexistence of BCJ relations for higher-point scalar correlators in Mellin space~\cite{Alday:2022lkk}. It will also potentially shed new light on AdS double-copy relations~\cite{Zhou:2021gnu} beyond four points.

We also hope to generalize our discussion to correlators involving fermions, as was also explored in \cite{Faller:2017hyt}. This will complete people's understanding of the Mellin formalism for arbitrary conformal correlators, and provide the necessary foundation to manifest supersymmetry in Mellin space. The definition would certainly depend on the spacetime dimension $d$. The most interesting case would be $d=4$, which relates to a series of important holographic models in $\mathrm{AdS}_5$, including the famous $\mathrm{AdS}_5 \times \mathrm{S}^5$ type IIB supergravity. Both the Mellin formalism and supersymmetry play important roles in these models, and it would be nice to find a way to combine these two. 

Finally, let us comment on the relation of operators with different spins in our formalism. On the most basic level, the possible kinematic dependence of spinning correlators can be classified using group-theoretical methods, and the building blocks at different values of the spin can be related by weight-shifting operators~\cite{Karateev:2017jgd}. These may help clarify the role of $\hat{\mathbb N}$ in the factorization formula, which seems to ``lower'' the spin of the exchanged operator. On a higher and more interesting level, it turns out that mathematically we can relate the $N$-point spinning Mellin amplitude to a $2N$-point ``scalar correlator'' $H$, where we no longer view these scalars as auxiliary but as actual. Using the pole-pinching mechanism reviewed in subsection~\ref{subsec:pinch}, we have\footnote{We make two remarks regarding this relation. First, we should define $H$ for generic $X_p,X_{p'}$ and only impose $X_{p'}\cdot X_p=0$ after taking the residues $\Delta_{p'}=-J_p$. Second, the operation $\mathop{\rm Res'}$ takes $(N-3)$ residues and then evaluates the remaining 3 at $\Delta_{p'}=-J_p$. The result is independent of how we split $\{1',\cdots,N'\}$ into $(N-3)$ and 3 variables.}
\begin{subequations}\label{eq:analyticity}
\begin{gather}
    G(X_1,X_{1'},\cdots,X_N,X_{N'})=\mathop{\rm Res'}_{\substack{\Delta_{1'}=-J_1\\\cdots\\\Delta_{N'}=-J_N}} H(X_1,X_{1'},\cdots,X_N,X_{N'})\Bigg|_{X_{p'}\cdot X_p=0},\\
    H(X_1,X_{1'},\cdots,X_N,X_{N'})=\int[{\rm d}\gamma]\,\mathcal M(\gamma)\prod_{\substack{u,v\in\mathcal S\\u\prec v}}\frac{\Gamma(\gamma_{uv})}{(-2X_u\cdot X_v)^{\gamma_{uv}}},\\
    \forall v\in\mathcal S=\{1,1',\cdots,N,N'\}:\quad\sum_{\substack{u\in\mathcal S\\u\neq v}}\gamma_{uv}=\Delta_v.
\end{gather}
\end{subequations}
In other words, integer-spin operators seem to arise as residues of a family of continuous-spin operators constructed out of pairs of null-separated scalars. This looks surprisingly similar to the properties of light-ray operators~\cite{Kravchuk:2018htv}, which suggests the exciting possibility of studying spin-analyticity~\cite{Caron-Huot:2017vep} or energy correlators~\cite{Moult:2025nhu} in the spinning Mellin formalism.

\begin{CJK*}{UTF8}{}
\CJKfamily{gbsn}
\acknowledgments
It is our pleasure to thank Vasco Gon\c{c}alves, Song He (何颂), Xiang Li (李想), Wen-Jie Ma (马文杰) for fruitful discussions, and especially Qu Cao (曹趣) and Ellis Ye Yuan (袁野) for collaboration at early stages of this project. ZH is supported by the National Science Foundation of China under Grant No.~12175197 and Grant No.~12347103. YT is supported by the National Science Foundation of China under Grant No.~12225510 and Grant No.~12447101, and by the New Cornerstone Science Foundation.
\end{CJK*}

\appendix{}

\section{Constructing canonical Mellin amplitudes}\label{app:fac}

As discussed in sections~\ref{subsec:discrete} and~\ref{subsec:singlespin}, only the values of a Mellin amplitude at the lattice points satisfying the following constraints are meaningful,
\begin{equation}\label{eq:spinsupp}
    \sum_{\substack{q=1\\q\neq p}}^N
    \big(\eta_{p'q}+\eta_{p'q'}\big)=J_p,
    \qquad
    \eta_{p'q},\eta_{p'q'}\in\mathbb Z_{\geq0}.
\end{equation}
We prefer to write each pole contribution in the form
\begin{equation}\label{eq:canpole}
    \frac{p(\gamma,\eta)}{\gamma_{LR}-\tau-2m},
\end{equation}
with the pole structure predicted by the OPE limit (see section~\ref{sec:ope2}) and a numerator that is multi-homogeneous in factorial powers of $\eta$ (we count the degree of $(\eta)^{\underline{n}}$ as $n$). Each term should have  degree $J_p$ in label $p'$,
\begin{equation}\label{eq:facdeg}
    \sum_{\substack{q=1\\q\neq p}}^N
    \big(n_{p'q}+n_{p'q'}\big)=J_p.
\end{equation}
Especially, the variable $\eta_{p'q'}$ contributes to the degrees of both $p'$ and $q'$.

Given a rational Mellin amplitude, we can always turn its denominator into the above form. The problem then reduces to turning a polynomial in $\eta$ into a multi-homogeneous polynomial in factorial powers.

\subsection{Single-spin case}

For simplicity, we first illustrate the construction with an $(N+1)$-point correlator of $N$ scalar operators and one symmetric traceless operator of spin $J$,
\begin{equation*}
    \big\langle \mathcal O_1(X_1)\cdots\mathcal O_N(X_N)
    \mathcal O_L(X_L,Z_L)\big\rangle.
\end{equation*}
Here $i=1,\ldots,N$ labels the scalar insertions, $L$ labels the spinning insertion, and $L'$ labels its polarization vector $Z_L$. The discrete Mellin variables $\eta_{iL'}$ are associated with the products $2X_i\cdot Z_L$. There is then only one spin constraint,
\begin{equation}\label{eq:singlesupp}
    \sum_{i=1}^N\eta_{iL'}=J,
    \qquad \eta_{iL'}\in\mathbb Z_{\geq0}.
\end{equation}

To illustrate the strategy, consider first an ordinary polynomial of total degree at most $J$,
\begin{equation}\label{eq:poly}
    p(\eta)=\sum_{\substack{a_1,\ldots,a_N\geq0\\\sum_i a_i\leq J}}
    c_{a_1\cdots a_N}\eta_{1L'}^{a_1}\cdots\eta_{NL'}^{a_N}.
\end{equation}
How can we turn it into a homogeneous polynomial of degree $J$? For $J>0$, we simply multiply each term by a sufficient power of $\sum_i\eta_{iL'}/J$, which equals one on the support,
\begin{equation}\label{eq:hom}
    \begin{aligned}
    p(\eta)
    &=\sum_{\substack{a_1,\ldots,a_N\geq0\\\sum_i a_i\leq J}}
    c_{a_1\cdots a_N}\eta_{1L'}^{a_1}\cdots\eta_{NL'}^{a_N}
    \times 1^{J-\sum_i a_i}\\
    &=\sum_{\substack{a_1,\ldots,a_N\geq0\\\sum_i a_i\leq J}}
    c_{a_1\cdots a_N}\eta_{1L'}^{a_1}\cdots\eta_{NL'}^{a_N}
    \left(\frac{\sum_i\eta_{iL'}}{J}\right)^{J-\sum_i a_i}.
    \end{aligned}
\end{equation}
The result is then a homogeneous polynomial of degree $J$.

The same strategy applies to the factorial powers. First, we rewrite the polynomial as
\begin{equation}\label{eq:facpoly}
    p(\eta)=\sum_{a_1,\ldots,a_N\geq0}
    \tilde c_{a_1\cdots a_N}
    (\eta_{1L'})^{\underline{a_1}}\cdots(\eta_{NL'})^{\underline{a_N}},
\end{equation}
using the Stirling formula
\begin{equation}\label{eq:stirling}
    x^k=\sum_{n=0}^k B_{n,k}(x)^{\underline n},\qquad
    B_{n,k}=\frac{1}{n!}\sum_{r=0}^n (-1)^{n-r}\binom nr r^k.
\end{equation}
For each term with $\sum_i a_i<J$, we can increase its factorial degree by one by multiplying by $\sum_i(\eta_{iL'}-a_i)/(J-\sum_i a_i)$, which equals one on the support,
\begin{equation}\label{eq:raise}
    \begin{aligned}
    &\tilde c_{a_1\cdots a_N}
    (\eta_{1L'})^{\underline{a_1}}\cdots(\eta_{NL'})^{\underline{a_N}}
    \;\frac{\sum_i(\eta_{iL'}-a_i)}{J-\sum_i a_i}\\
    &\qquad=\frac{\tilde c_{a_1\cdots a_N}}{J-\sum_i a_i}
    \sum_i(\eta_{iL'})^{\underline{a_i+1}}
    \prod_{j\neq i}(\eta_{jL'})^{\underline{a_j}}.
    \end{aligned}
\end{equation}
Here we have used $(\eta_{iL'})^{\underline{a_i}}(\eta_{iL'}-a_i)=(\eta_{iL'})^{\underline{a_i+1}}$. Repeating this step gives a homogeneous polynomial of factorial degree $J$.

\subsection{Multi-spin case}

For several spinning operators, we use the same method to raise the degree in each label $p'$ to $J_p$. Raising a factorial power of $\eta_{p'q'}$ increases the degrees of both $p'$ and $q'$. In this process, one might generate terms with degrees higher than the corresponding spins. These terms vanish on the lattice and can be simply dropped. 

Let us illustrate the multi-spin case with a simple example. Consider the four spin-one example in section~\ref{subsec:discrete}, involving only products $Z_p\cdot Z_q$. The discrete variables $\eta_{p'q'}$ satisfy
\begin{equation}\label{eq:fourspin}
    \sum_{\substack{q=1\\q\neq p}}^4\eta_{p'q'}=1, \qquad \eta_{p'q'}\in\mathbb Z_{\geq0}.
\end{equation}
Starting with the constant polynomial $1$, we first raise the degree in label $1'$,
\begin{equation}\label{eq:raise1}
    1=\eta_{1'2'}+\eta_{1'3'}+\eta_{1'4'}.
\end{equation}
The first term already has degree one in labels $1'$ and $2'$. Raising its degree in $3'$ gives
\begin{equation}\label{eq:raise2}
    \begin{aligned}
    \eta_{1'2'}=\eta_{1'2'}(\eta_{1'3'}+\eta_{2'3'}+\eta_{3'4'})=\eta_{1'2'}\eta_{3'4'}.
    \end{aligned}
\end{equation}
The products $\eta_{1'2'}\eta_{1'3'}$ and $\eta_{1'2'}\eta_{2'3'}$ vanish by the spin-one constraints. The remaining term has degree one in all four labels. Treating the other two terms in the same way, we obtain
\begin{equation}\label{eq:facex}
    1=
    \eta_{1'2'}\eta_{3'4'}
    +\eta_{1'3'}\eta_{2'4'}
    +\eta_{1'4'}\eta_{2'3'}
    \qquad\text{(on the physical lattice points)}.
\end{equation}
This is the desired homogeneous expression with each term having degree one in all four labels, as $(\eta)^{\underline1}=\eta$.

\bibliographystyle{utphys}
\bibliography{refs}

\end{document}